\documentclass[a4paper,fleqn,usenatbib]{mnras}

\usepackage{amsmath}	
\usepackage{amssymb}	
\usepackage{newtxtt,newtxmath}

\usepackage[T1]{fontenc}
\usepackage{ae,aecompl}

\usepackage{graphicx}	
\usepackage{epstopdf}
\usepackage{subcaption}
\usepackage{multicol}
\usepackage{multirow}
\usepackage{float}
\usepackage[usenames,dvipsnames,svgnames,table]{xcolor}
\usepackage{enumitem}
\usepackage{bm}
\usepackage{tablefootnote}

\usepackage[noabbrev]{cleveref}

\crefname{equation}{}{}
\crefname{figure}{}{}

\newcommand{\lya}{Ly$\alpha$}
\newcommand{\ha}{H$\alpha$}
\newcommand{\hb}{H$\beta$}
\newcommand{\oiii}{[O{\sc iii}]}
\newcommand{\oii}{[O{\sc ii}]}
\newcommand{\nii}{[N{\sc ii}]}

\newcommand{\msol}{M$_\odot$}

\newcommand{\aha}{$A_\textrm{\ha}$}
\newcommand{\pstar}{$\phi^\star$}
\newcommand{\lstar}{$L^\star$}

\title[LAGER: Luminosity Functions]{A large, deep 3 deg$^2$ survey of \ha, \oiii, and \oii~emitters from LAGER: constraining luminosity functions}
\author[Khostovan et al.]{A.~A.~Khostovan$^{1}$\thanks{NASA Postdoctoral Program Fellow}\thanks{E-mail:
akhostov@gmail.com}, S.~Malhotra$^{1,2}$, J.~E.~Rhoads$^{1,2}$, C.~Jiang$^{3}$, J.~Wang$^{4}$, I.~Wold$^{1}$, \newauthor Z.-Y.~Zheng$^{3}$, L.~F.~Barrientos$^{6}$, A~.Coughlin$^{2,5}$, S.~Harish$^{2}$, W.~Hu$^{4}$, L.~Infante$^{6,7,8}$, \newauthor L.~A.~Perez$^{2}$, J.~Pharo$^{2}$, F.~Valdes$^{9}$, A.~R.~Walker$^{10}$, H.~Yang$^{7}$\\ 
$^{1}$Astrophysics Division, NASA Goddard Space Flight Center, Greenbelt, MD 20771, United States of America\\
$^{2}$School of Earth and Space Exploration, Arizona State University, Tempe, AZ 85287, United States of America\\
$^{3}$CAS Key Laboratory for Research in Galaxies and Cosmology, Shanghai Astronomical Observatory, Shanghai 200030, China\\
$^{4}$CAS Key Laboratory for Research in Galaxies and Cosmology, University of Science and Technology of China,
Hefei, Anhui 230026\\
$^{5}$Chandler-Gilbert Community College, 2626 East Pecos Road, Chandler, AZ 85225-2499\\
$^{6}$Instituto de Astrof\'isica, Facultad de F\'isica, Pontificia Universidad Cat\'olica de Chile, Santiago, Chile\\
$^{7}$Las Campanas Observatory, Observatories of the Carnegie Institution of Washington, La Serena, Chile\\
$^{8}$N\'ucleo de Astronom\'ia, Universidad Diego Portales, Santiago, Chile\\
$^{9}$National Optical Astronomy Observatory, 950 N. Cherry Avenue, Tucson, AZ 85719, United States of America\\
$^{10}$Cerro Tololo Inter-American Observatory, National Optical Astronomy Observatory, Casilla 603, La Serena, Chile}

\date{}

\pubyear{2019}
\begin{document}

\label{firstpage}
\pagerange{\pageref{firstpage}--\pageref{lastpage}}
\maketitle

\begin{abstract} 
We present our measurements of the \ha, \oiii, and \oii~luminosity functions as part of the Lyman Alpha Galaxies at Epoch of Reionization (LAGER) survey using our samples of 1577 $z = 0.47$ \ha-, 3933 $z = 0.93$ \oiii-, and 5367 $z = 1.59$ \oii-selected emission line galaxies in a 3 deg$^2$ single, CTIO/Blanco DECam pointing of the COSMOS field. Our observations reach 5$\sigma$ depths of $8.2 \times 10^{-18}$ erg s$^{-1}$ cm$^{-2}$ and comoving volumes of $(1-7) \times 10^{5}$ Mpc$^3$ making our survey one of the deepest narrowband surveys. We select our emission line galaxies via spectroscopic confirmation, photometric redshifts, and color-color selections. We measure the observed luminosity functions for each sample and find best-fits of $\phi^\star = 10^{-3.16^{+0.09}_{-0.09}}$ Mpc$^{-3}$ and $L^\star = 10^{41.72^{+0.09}_{-0.09}}$ erg s$^{-1}$ for \ha, $\phi^\star = 10^{-2.16^{+0.10}_{-0.12}}$ Mpc$^{-3}$ and $L^\star = 10^{41.38^{+0.07}_{-0.06}}$ erg s$^{-1}$ for \oiii, and $\phi^\star = 10^{-1.97^{+0.07}_{-0.07}}$ Mpc$^{-3}$ and $L^\star = 10^{41.66^{+0.03}_{-0.03}}$ erg s$^{-1}$ for \oii, with $\alpha$ fixed to $-1.75$, $-1.6$, and $-1.3$, respectively. An excess of bright $> 10^{42}$ erg s$^{-1}$ \oiii~emitters is observed and may be due to AGN contamination. Corrections for dust attenuation are applied assuming \aha$ = 1$ mag. We also design our own empirical rest-frame $g - r$ calibration using SDSS DR12 data, test it against our $z = 0.47$ \ha~emitters with $z$COSMOS $1$D spectra, and calibrate it for $(g - r)$ between $-0.8$ and $1.3$ mag. Dust and AGN-corrected star formation rate densities (SFRDs) are measured as $\log_{10} \rho_\textrm{SFR}/(\textrm{M}_\odot\ \textrm{yr}^{-1}\ \textrm{Mpc}^{-3}) = -1.63\pm0.04$, $-1.07\pm0.06$, and $-0.90\pm0.10$ for \ha, \oiii, and \oii, respectively. We find our \oiii~and \oii~samples fully trace cosmic star formation activity at their respective redshifts in comparison to multi-wavelength SFRDs, while the \ha~sample traces $\sim 70$ percent of the total $z = 0.47$ SFRD.

\end{abstract}


\section{Introduction}
Cosmic star formation is a fundamental property of galaxy formation and evolution physics as imprinted in it are all the physical, stochastic processes associated with star-formation that occurred at a given epoch. Observationally constraining this important property is necessary for future work in understanding the underlying physics. 

The current, overall consensus is that galaxies, in general, became rapidly active during the first 2 - 3 Gyrs with a peak around $z \sim 2 - 3$ and have slowly declined in star formation activity over the past 10 - 11 Gyrs till the present-day (e.g., \citealt{Hopkins2006,Madau2014,Khostovan2015}). Despite the wealth of progress made, the cosmic star formation history is still plagued with scatter, which arises from different star formation calibrations, dust corrections, sample variances, selection biases, and various other factors. Consistent measurements are then needed to properly constrain the cosmic star formation history, which relies on fully understanding the statistical properties of star-forming galaxies. 

The luminosity function of star-forming galaxies is one such statistical property that traces the distribution of galaxies as dependent on their continuum or emission line luminosity (tracers of star-formation activity). There exists a plethora of continuum-based luminosity functions based on UV (e.g., \citealt{Reddy2009,Cucciati2012,Bouwens2015}) and IR (e.g., \citealt{Magnelli2011,Magnelli2013,Gruppioni2013}), which cover a wide range of cosmic time. Although these studies have greatly enhanced our understanding of galaxy evolution, they rely heavily on having accurate photometric redshifts. Furthermore, continuum-based studies select their samples within photo-$z$ bins which, in terms of cosmic time, can incorporate 100s Myr to a few Gyrs depending on the central redshift. This raises concern on cosmic evolutionary effects embedded within the corresponding measurements. What we then require are samples with accurate redshifts within a narrow range in cosmic time to mitigate internal evolutionary effects.

Narrowband surveys are useful in addressing this issue as they observe `active' galaxies based on their nebular emission lines, which allows for accurate redshift measurements assuming correct emission line identification. Depending on the design, narrowband filters can have a FWHM of a few 10s to $\sim 100$\AA~that corresponds to redshift windows of $\sim 0.01 - 0.05$, which allows for thin slices in cosmic time. Furthermore, nebular emission lines such as \ha, \oiii, and \oii~trace massive, bright $O$ and $B$ type stars corresponding to star formation timescales of $\sim 10$ Myr, such that narrowband surveys can trace the instantaneous star formation activity at a given epoch.

Previous narrowband surveys have primarily focused on \ha~as it is a well-known tracer of star formation activity and less prone to dust attenuation (e.g., \citealt{Gallego1995,Tresse1998,Fujita2003,Hippelein2003,Ly2007,Morioka2008,Shioya2008,Ly2011,Tadaki2011,Sobral2013,Sobral2015,Stroe2015,Matthee2017,Coughlin2018,Hayashi2018}). Other emission lines, such as \oiii~(e.g., \citealt{Hippelein2003,Ly2007,Khostovan2015,Sobral2015,Matthee2017,Hayashi2018}) and \oii~(e.g., \citealt{Ly2007,Takahashi2007,Bayliss2011,Bayliss2012,Sobral2012,Khostovan2015,Sobral2015,Matthee2017,Hayashi2018}), have also been observed with narrowband surveys. The majority of narrowband surveys observe \ha, \oiii, and \oii~up to $z \sim 0.5$, $0.9$, and $1.6$, respectively, as each of the respective lines fall into the near-IR at higher redshifts. {\it HST} grism surveys have helped extend this window to higher redshifts, such as with WISP \citep{Colbert2013}, PEARS \citep{Pirzkal2013}, and FIGS \citep{Pirzkal2018}, but cover wider redshift ranges per sample in comparison to narrowband surveys.

Recent advances in near-IR detector technology have allowed for narrowband studies to observe the three strong nebular emission lines up to higher redshifts. Such studies include the Deep and Wide Narrowband (DAWN) survey, which used a custom-made NB1066 filter to observe $z = 0.62$ \ha~emitters (\citealt{Coughlin2018}, Harish et al., submitted). The 2 deg$^2$ High-$z$ Emission Line Survey (HiZELS) used four narrowband filters in $z$, $J$, $H$, and $K$ to observe \ha~emitters up to $z \sim 2$ \citep{Sobral2013}, as well as \oiii~and \oii~emitters up to $z \sim 3$ and $5$, respectively \citep{Khostovan2015}. 
	
Past measurements of the emission line luminosity functions were based on small area surveys (< 1 deg$^2$; e.g., \citealt{Ly2007,Tadaki2011,Bayliss2011,Bayliss2012,Sobral2012}). Although such surveys can better measure the faint-end slope, they can not constrain the bright-end of the luminosity function given the low volumes as their number densities are significantly smaller. Furthermore, cosmic variance effects also raise uncertainties on luminosity function properties. \citet{Sobral2015} used the large 10 deg$^2$ CF-HiZELS samples to assess the effects of cosmic variance on the characteristic line luminosity, \lstar, and number density, \pstar. They reported $z = 0.84$ \ha~samples covering $< 10^4$ Mpc$^3$ ($< 1$ deg$^2$) are $>50\%$ uncertain in \lstar~and \pstar~due to their limited volume alone. \citet{Stroe2015} and \citet{Shioya2008} both observed $z = 0.24$ \ha~emitters with a survey coverage of 26 deg$^2$ and 1.5 deg$^2$, respectively. Assuming the same faint-end slope of $\alpha = -1.35$ as measured by \citet{Shioya2008}, \citet{Stroe2015} reported $L^\star = 10^{41.71\pm0.02}$ erg s$^{-1}$ in comparison to $L^\star = 10^{41.54^{+0.38}_{-0.29}}$ erg s$^{-1}$ measured by \citet{Shioya2008}. The $0.24$ deg$^2$ Subaru Deep Field (SDF) $z = 0.24$ measurements of \citet{Ly2007} report $L^\star  =  10^{41.25\pm0.34}$ erg s$^{-1}$, although they find $\alpha = -1.70\pm0.10$, much steeper than that reported in \citet{Shioya2008}. This highlights the importance of wide-field narrowband observations to set proper constrains on the bright-end, while also the need for deep observations to constrain the faint-end of the luminosity functions.
 
\citet{Sobral2013} reported the \ha~luminosity functions evolving as $\log_{10} L^\star(z) = 0.45z + 41.87$ in characteristic line luminosity, $L^\star$, up to $z \sim 2$ (assuming \aha$ = 1$ mag). Their measurements also show a faint-end slope consistent with $\alpha = -1.6$ across their redshift range. Larger narrowband surveys such as the 10 deg$^2$ CF-HiZELS \citep{Sobral2015} and the 16 deg$^2$ HyperSuprimeCam (HSC) survey \citep{Hayashi2018} report $L^\star(z)$ consistent with the evolution measured in \citet{Sobral2013}. The \oiii~and \oii~LFs are found to evolve considerably up to $z \sim 3$ and $5$, respectively, with \citet{Khostovan2015} reporting a 2 and 3 dex increase in $L^\star$ for \oiii~and \oii~LFs, respectively. 

Past results of the luminosity functions show the underlying need for wide-field, deep narrowband surveys that can cover a wide range in line luminosities that would constrain the bright-end of the luminosity function, as well as the faint-end. Constraining the bright-end is of significant importance for planning future wide-field surveys with {\it WFIRST} and {\it Euclid} that rely on accurate number count predictions. Observations of the bright-end also allow us to study rare, extreme `active' galaxies and understand the underlying physics that drives the emission line production. Several studies have attempted to empirically model the \ha~luminosity function evolution for number count predictions of future surveys (e.g., \citealt{Pozzetti2016}), but are limited to primarily small area ($\sim 0.01 - 0.1$ deg$^2$) surveys and a handful of large ($> 1$ deg$^2$) surveys for constraining their models. This highlights the need for large, narrowband surveys to investigate the statistical properties of emission line galaxies.

In this paper, we present the first results of \ha, \oiii, and \oii~emitters in the COSMOS field as part of the Lyman Alpha Galaxies in the Epoch of Reionization (LAGER) survey. Using a customized NB964 filter with the DECam instrument installed on CTIO/Blanco, we measure the luminosity functions for each respective emission line in a 3 deg$^2$ pointing. The wide area and comparable depth to HiZELS allows us to constrain the bright-end of the luminosity function as well as the faint-end. Upon completion, LAGER will have observed a total of 8 fields resulting in a complete survey coverage of 24 deg$^2$.

The paper is organized as follows: in \S \ref{sec:samples}, we describe our observations, data reduction, object identification and cataloging, methodology of sample selection, and investigation of the contaminants from misidentified lines and active galactic nuclei. In \S \ref{sec:methodology}, we present our methodology in measuring the luminosity functions where we take into account \nii~contamination in our \ha~samples, measure the completeness of each sample, calculate the filter profile corrections, describe our dust correction methods, and measure the luminosity functions using the $V_\textrm{max}$ estimator. In \S \ref{sec:LFs}, we present our luminosity function results for each of our emission line galaxy samples. In \S \ref{sec:SFRDs}, we use our luminosity functions to assess the cosmic star formation rate densities. In \S \ref{sec:conclusion}, we highlight our main results and present final remarks regarding the complete LAGER survey.

We assume a $\Lambda$CDM cosmology with $H_0$ = 70 km s$^{-1}$ Mpc$^{-1}$, $\Omega_m = 0.3$, and $\Omega_\Lambda = 0.7$. A Salpeter initial mass function (IMF) is assumed for all related measurements. All magnitudes shown are based on the AB system unless otherwise stated.

\begin{figure}
	\centering
	\includegraphics[width=\columnwidth]{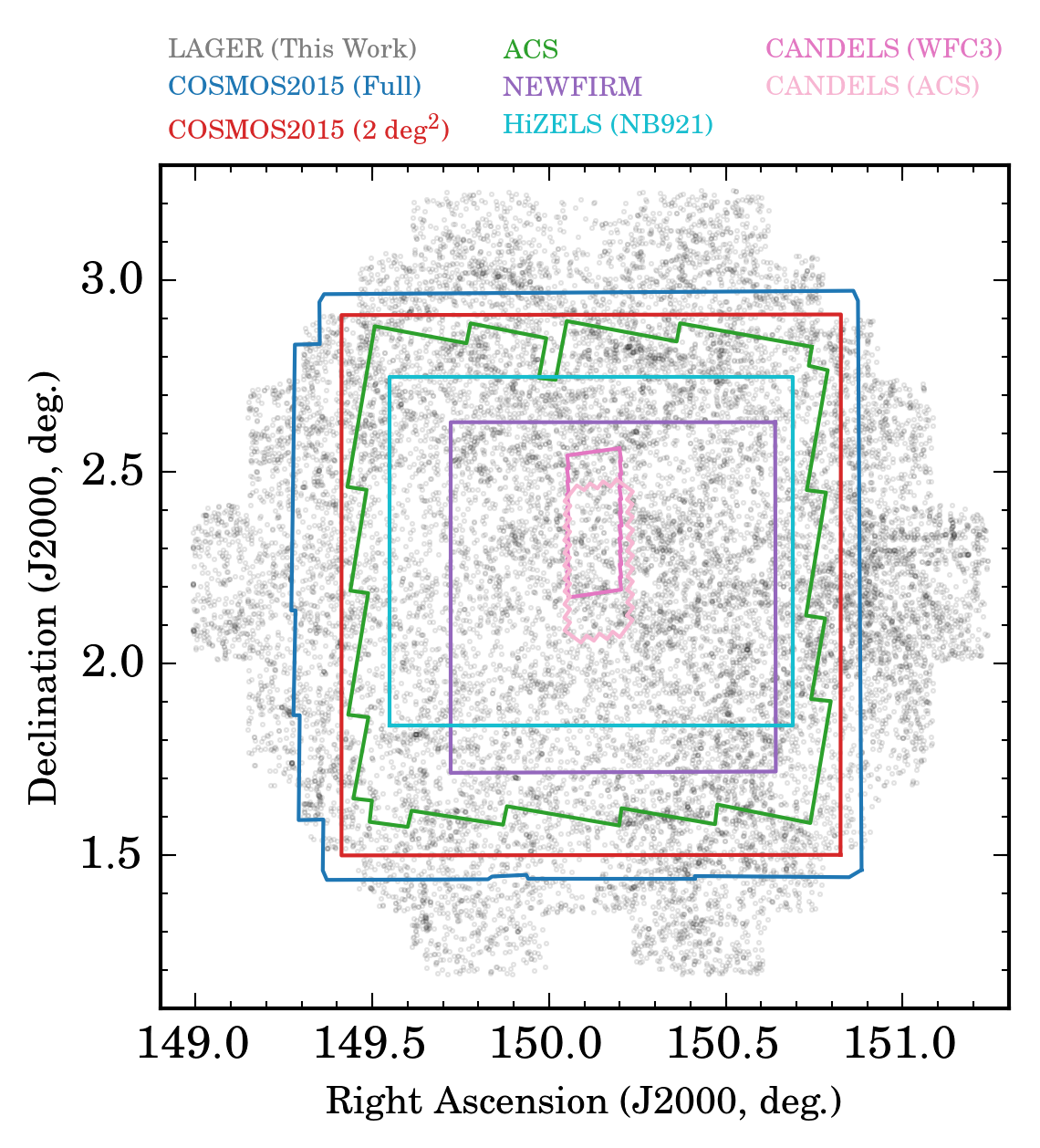}
	\caption{The LAGER Survey footprint, shown in {\it grey} in comparison to several surveys with coverage of the COSMOS field. The full 2.4 deg$^2$ $zYJHK_s$ stacked image coverage of the COSMOS2015 survey of \citet{Laigle2016} is shown in {\it blue} and the corresponding 2 deg$^2$ coverage in {\it red}. The 1.64 deg$^2$ {\it HST}/ACS F814W coverage is shown in {\it green} \citep{Koekemoer2007}. The NEWFIRM  Medium-Band Survey (NMBS) coverage is shown in {\it purple} \citep{Whitaker2011} and the HiZELS NB921 coverage is shown in {\it sky blue} \citep{Sobral2013}. We also show the CANDELS {\it HST}/ACS and WFC3 coverage in {\it light} and {\it dark pink}, respectively \citep{Grogin2011}. Our survey covers an area wider than COSMOS2015 and the ancillary photometry within that catalog, such that we restrict our selection and subsequent analyses to the 2.4 deg$^2$ coverage of the full COSMOS2015 detection area.}
	\label{fig:LAGER_footprint}
\end{figure}

\section{LAGER}
\label{sec:samples}

\subsection{Observations \& Data Reduction}

\begin{figure}
	\centering
	\includegraphics[width=\columnwidth]{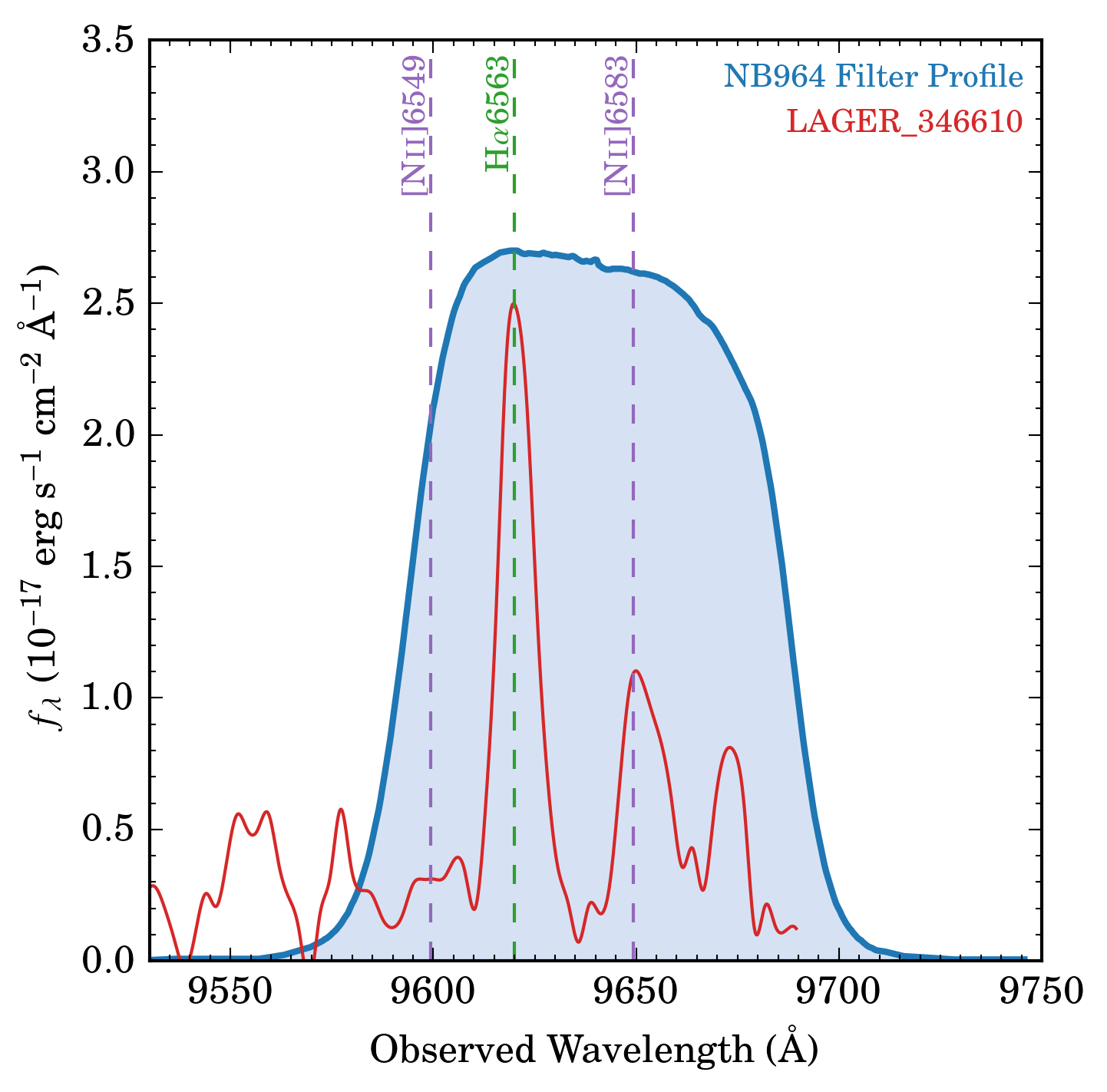}
	\caption{Our NB964 narrowband filter profile designed by \citet{Zheng2019} with central wavelength of 9640\AA~and full-width-half-max (FWHM) of 92\AA. The filter is not a perfect top-hat, such that filter profile corrections are needed when assessing the luminosity function. Overlaid is the 1D flux-calibrated $z$COSMOS spectra of one of our $z = 0.47$ \ha-selected emitters. The lack of spectra redwards of 9700\AA~is due to wavelength limitations of VLT/VIMOS. Both \nii~lines can be observed along with the \ha~line, such that not correcting for \nii~contamination can result in overestimation of the observed \ha~line luminosity.}
	\label{fig:NB_profile}
\end{figure}

The Lyman Alpha Galaxies in the Epoch of Reionization (LAGER) survey, shown in Figure \ref{fig:LAGER_footprint}, uses a custom narrowband filter (NB964; \citealt{Zheng2019}) with a central wavelength of 9640\AA~and FWHM of 92\AA, with the filter profile shown in Figure \ref{fig:NB_profile}. The filter was specifically designed to avoid atmospheric absorption and OH emission lines. Observations were done using the narrowband filter and the Dark Energy Camera (DECam) installed on the Cerro Tololo Inter-American Observatory (CTIO) 4-m Blanco Telescope. DECam is a wide-field camera that covers 3 deg$^2$ per pointing using 62 science CCDs with a resolution of 0.263$''$/pixel. Observations of the COSMOS field were done in between December 2015 and 2017 with a total exposure time of 47.25 hours. Although the main science case for LAGER was \lya-related science \citep{Hu2017,Zheng2017,Hu2019,Yang2019}, we extend the deep, wide-field capabilities of this survey to investigate the foreground emission line galaxies (\ha, \oiii, and \oii-selected emitters). We use DECam-$z$ photometry as the broadband (BB) counterpart to our NB964 data. All the DECam-$z$ images are available via the National Optical Astronomy Observatory (NOAO) Science Archive.

We refer the reader to \citet{Hu2019} for details on the data reduction and source extraction. Briefly, the data was reduced and calibrated using the DECam community pipeline \citep{Valdes2014}. Individual DECam frames were then stacked via a customized pipeline described in \citet{Hu2019}, which creates a $\sim 3$ deg$^2$ stacked science mosaic with a corresponding weight map that takes into account the point spread function, exposure time, and atmospheric transmission.

Source extraction was done using {\sc SExtractor} \citep{Bertin1996} in dual-imaging mode (NB964 and DECam $z$; associated broadband). The DECam $z$ zeropoint was calibrated by using bright stars in the field and the associated Subaru SuprimeCam photometry of the $K_s-$selected UltraVISTA DR1 catalog \citep{Muzzin2013}. The narrowband zeropoint was calibrated by taking bright $A-$ and $B$-type stars in each field, fitting their spectral energy distributions, and then convolving their spectra with the NB964 filter profile to measure the auto magnitude per source. The measured zeropoints are 28.77 and 32.37 mag for the NB964 and DECam $z$ band, respectively, and all source magnitudes further mentioned in this paper are based on 2$''$ apertures. The $5\sigma$ limiting AB magnitudes are 25.45 and 25.84 mag for NB964 and DECam $z$, respectively.

\subsection{Sample Selection}
\label{sec:sample_selection}

\subsubsection{Emission Line Galaxy Candidate Selection}
\label{sec:ELG_selection}

\begin{figure}
	\centering
	\includegraphics[width=\columnwidth]{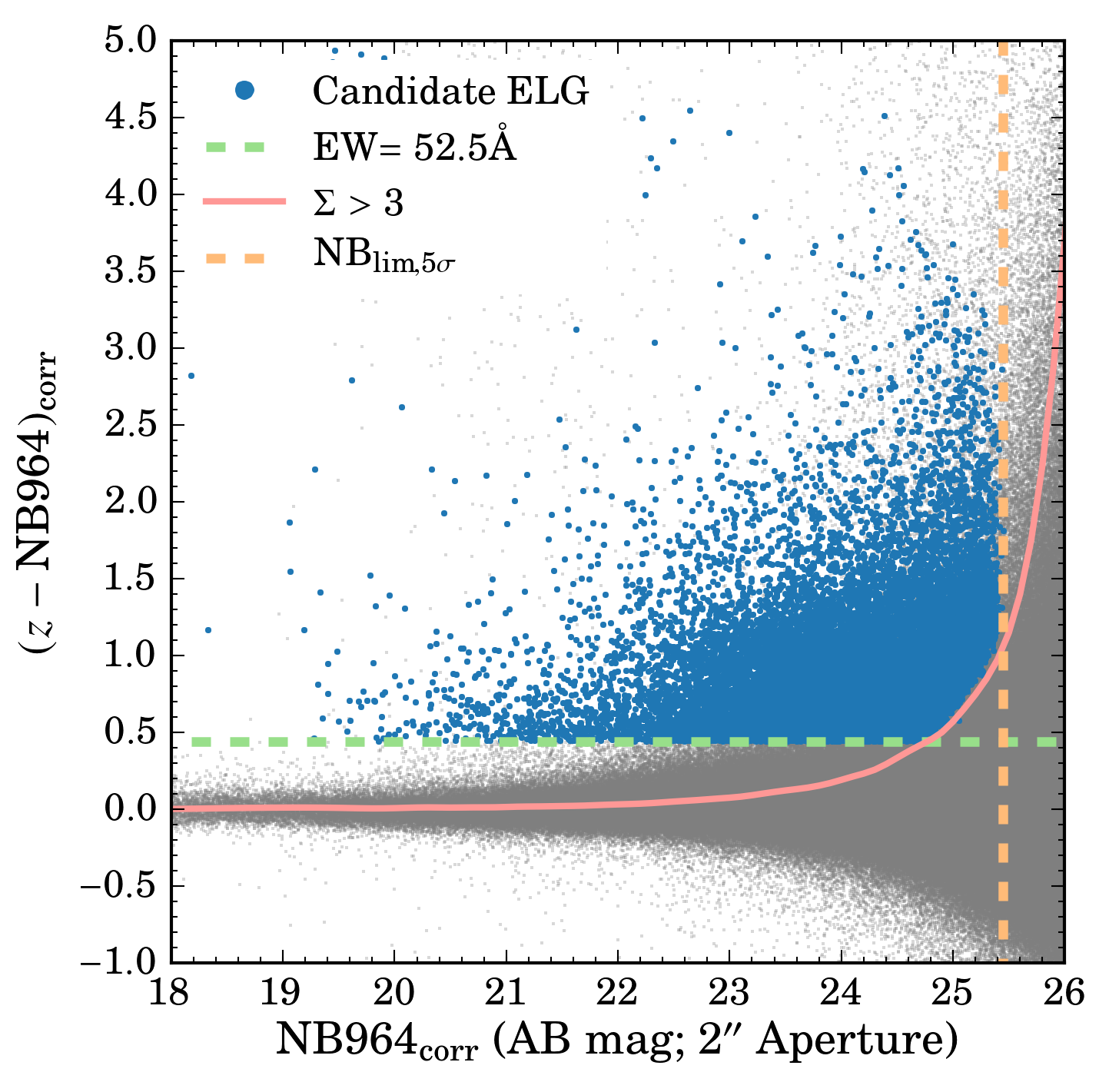}
	\caption{The nebular color excess and emission line galaxy candidate selection based on DECam NB964 and $z$-band photometry. An observed equivalent width cut of $>52.5$\AA~is placed, corresponding to $(z - \textrm{NB964})$ = 0.44 mag ($f^\textrm{NB}_\nu/f^z_\nu = 1.5$) that removes the sources with bright narrowband and continuum fluxes not associated with ELGs. A color significance $\Sigma > 3$ cut is applied to take into account photometric scatter at faint narrowband magnitudes and a $>5\sigma$ NB magnitude detection limit is placed to remove potential contamination due to noise. The color significance criteria also removes sources within the EW and 5 $\sigma$ NB magnitude limit selection range. These are sources for which their photometry, be it the NB or BB, are highly uncertain and do not satisfy our $\Sigma > 3$ selection. A total of 18,268 sources are selected and are shown as {\it blue circles}.}
	\label{fig:NB_excess}
\end{figure}

We select our emission line galaxy (ELG) candidates by looking at their nebular color excess (BB -- NB), where BB and NB are the broadband and narrowband filters, respectively. The selection assumes that (BB -- NB) $= 0$ mag in the case of no emission line within the narrowband filter. Our NB964 filter is $\sim 381$\AA~redwards from the center of the DECam $z$ filter, such that sources with strong continuum colours will not have a consistent (BB -- NB) $= 0$ mag. Not correcting for this factor results in a $+0.254$ mag median offset in the color excess. We correct for this effect by investigating the continuum-dominated sources (those that exhibit BB -- NB $\sim 0.25$ mag and within the $1\sigma$ scatter of 0.315 mag) and see if they exhibit strong ($z - Y$) colors, with $Y$ band photometry from Subaru HSC. We find no $z - Y$ dependency on the $z - $NB color, such that we correct all our NB magnitudes by making them 0.254 mag fainter. This then ensures a nebular color excess of 0 for the case where no emission line falls within the narrowband filter.

Figure \ref{fig:NB_excess} shows the nebular color-magnitude selection of our (ELG) candidates with the color correction applied. A 5$\sigma$ NB magnitude limit of 25.45 mag is applied along with an observed equivalent width (EW) of $52.5$\AA, corresponding to a $(\textrm{BB} - \textrm{NB}) = 0.44$ mag ($f^\textrm{NB}_\nu/f^\textrm{BB}_\nu$ = 1.5). The EW cut ensures that we are selecting candidates that include an emission line rather than strong continuum features. The 5$\sigma$ limit reduces the level of artificial sources that could contaminate our sample.

The last ELG selection applied was a color significance cut, $\Sigma$, which ensures that the narrowband excess of each emission line source is significant above a specific S/N cut (in our case above 3$\sigma$). This is described as:
\begin{eqnarray}
\Sigma = \frac{1 - 10^{-0.4(\textrm{BB} - \textrm{NB})}}{10^{\textrm{ZP} - \textrm{NB}} \sqrt{\sigma_{\textrm{BB}}^2 + \sigma_{\textrm{NB}}^2}}
\end{eqnarray}
where the NB magnitudes include the color corrections that take into account the narrowband and broadband filter offsets, such that the zeropoint (ZP) applied in measuring $\Sigma$ is that of the broadband filter (ZP $= 32.37$ mag). $\sigma_\textrm{BB}$ and $\sigma_\textrm{NB}$ are the photometric errors for the broadband and narrowband fluxes measured from {\sc Sextractor}, respectively. We apply a $\Sigma > 3$ cut, which removes sources with NB magnitudes of 25 mag to the 5$\sigma$ limit as shown in Figure \ref{fig:NB_excess}. The color significance cut ensures that sources are not selected as ELG candidates due to photometric scatter. In total, we select 18,268 ELG candidates.

\begin{table*}
	\caption{Criteria applied for the selection of \ha, \oiii, and \oii~emitters. We select sources in the order of spectroscopic confirmation, photometric redshifts, and color-colors. We show the range of spectroscopic and photometric redshifts used in the selection below. The advantage of using the color-color selection is to pick up sources that lack enough continuum measurements to measure reliable photometric redshifts. These, in principle, should be low-mass, high equivalent width systems that would otherwise be missed using conventional continuum-selected techniques. Note that we also include \hb~with the \oiii~selection. This is due to the issue that photometric redshifts and color-color selection can not reliably differentiate between the two emission lines.}
	\begin{tabular*}{\textwidth}{@{\extracolsep{\fill}}cccc}
		\hline
		Line	& 	Spec-$z$ Criteria	& Photo-$z$ Criteria	& Color-Color Criteria \\
		\hline
		\ha 	& 	 0.454 -- 0.480 & 0.40 -- 0.55 & $(B - V) > 0.12$ and $(z - J) < 0.8$ and \\
		& 					& 			   & $0.8(z - J) + 0.05 < (B - V) < 0.8(z - J) + 0.5$\\
		&					& 				& \\
		\oiii,\hb   &	0.911 -- 0.960, 0.969 -- 0.997 & 0.70 -- 1.10 & for all $(z - J) < 0.8$ and $(B - V) > 0.8(z - J) - 0.4$\\
		& & & $(B - V) < 0.12$ or $(B - V) < 0.8(z - J) + 0.05$ \\
		&					& 				& \\
		\oii 	& 1.568 -- 1.604 & 1.30 -- 1.90 & $(B - V) < 0.8(z - J) - 0.4$\\
		\hline
	\end{tabular*}	
	\label{table:criteria}
\end{table*}

\subsubsection{Spectroscopic Selection}
\label{sec:spec_sel}

A major advantage of the COSMOS field is the wealth of spectroscopic surveys and programs that have accumulated over the years, which we use to select confirmed \ha, \oiii, and \oii~emission line galaxies. We create a compilation of 51,116 spectroscopic redshifts drawn from various surveys and observations:  10K-DEIMOS \citep{Hasinger2018}, 3D-$HST$ \citep{Brammer2012,Momcheva2016}, VLT/FORS2 observations \citep{Comparat2015}, C3R2 \citep{Masters2017}, FMOS-COSMOS \citep{Silverman2015}, GEEC2 \citep{Balogh2014}, COSMOS-\oii~\citep{Kaasinen2017}, LEGA-C \citep{Straatman2018}, MOSDEF \citep{Kriek2015}, PRIMUS \citep{Coil2011,Cool2013}, MMT/Hectospec observations \citep{Prescott2006}, Magellan/IMACS observations \citep{Trump2009}, and $z$COSMOS \citep{Lilly2009}. 

Spectroscopic selection ranges are defined by based on the full range of the NB filter transmission curve from 9570 -- 9705 \AA. Although the majority of sources are found within the FWHM of the NB filter, there are still intrinsically bright line emitters in our sample located towards the wings of the filter profile that appear observationally faint. For completeness, we ensure that such sources are also selected.

In total, we select 222 \ha~emitters between $0.454 < z < 0.480$, 120 \oiii~emitters in the range of $0.911 < z < 0.960$ and 4 \hb~emitters between $0.969 < z < 0.997$, and 59 \oii~emitters in the range of $1.568 < z < 1.604$, resulting in a total of 405 spectroscopically confirmed ELGs. We do not consider quality flags in the selection of sources within the inner range of the NB profile as the nebular color excess selection brings added confirmation to the measured redshift.

The large number of spectroscopic confirmations and also the availability of their 1D spectra allows us to assess the level of contamination in our other selection methods, as well as test our assumptions in regards to emission line properties (e.g., dust calibrations; see \S\ref{sec:dust_calib}). We caution that the spectroscopic data from the literature is biased towards bright sources and is not necessarily based on an emission line detection within the NB profile (e.g., 3D-{\it HST} G141 observes $z = 1.59$ \ha~while the \oii~line for the same source falls within our NB964 filter). The typical line fluxes for our spectroscopic confirmed sources are $1.2 \times 10^{-16}$, $5.9 \times 10^{-17}$, and $4.6 \times 10^{-17}$ erg s$^{-1}$ cm$^{-2}$ for \ha, \oiii, and \oii, respectively, whereas our samples have a $5\sigma$ limiting line flux of $8.2 \times 10^{-18}$ erg s$^{-1}$ cm$^{-2}$. The typical observed continuum fluxes are $1.45 \times 10^{-18}$, $5.32 \times 10^{-19}$, and $3.65 \times 10^{-19}$ erg s$^{-1}$ cm$^{-2}$ \AA$^{-1}$ for our spectroscopically confirmed sources while the typical continuum flux for each full sample is $\sim 5.9$, $4.3$, and $3.1$ times fainter for \ha, \oiii, and \oii-selected sources, respectively. This highlights the selection bias of our spectroscopically confirmed samples towards bright continuum systems.

\begin{figure}
	\centering
	\includegraphics[width=\columnwidth]{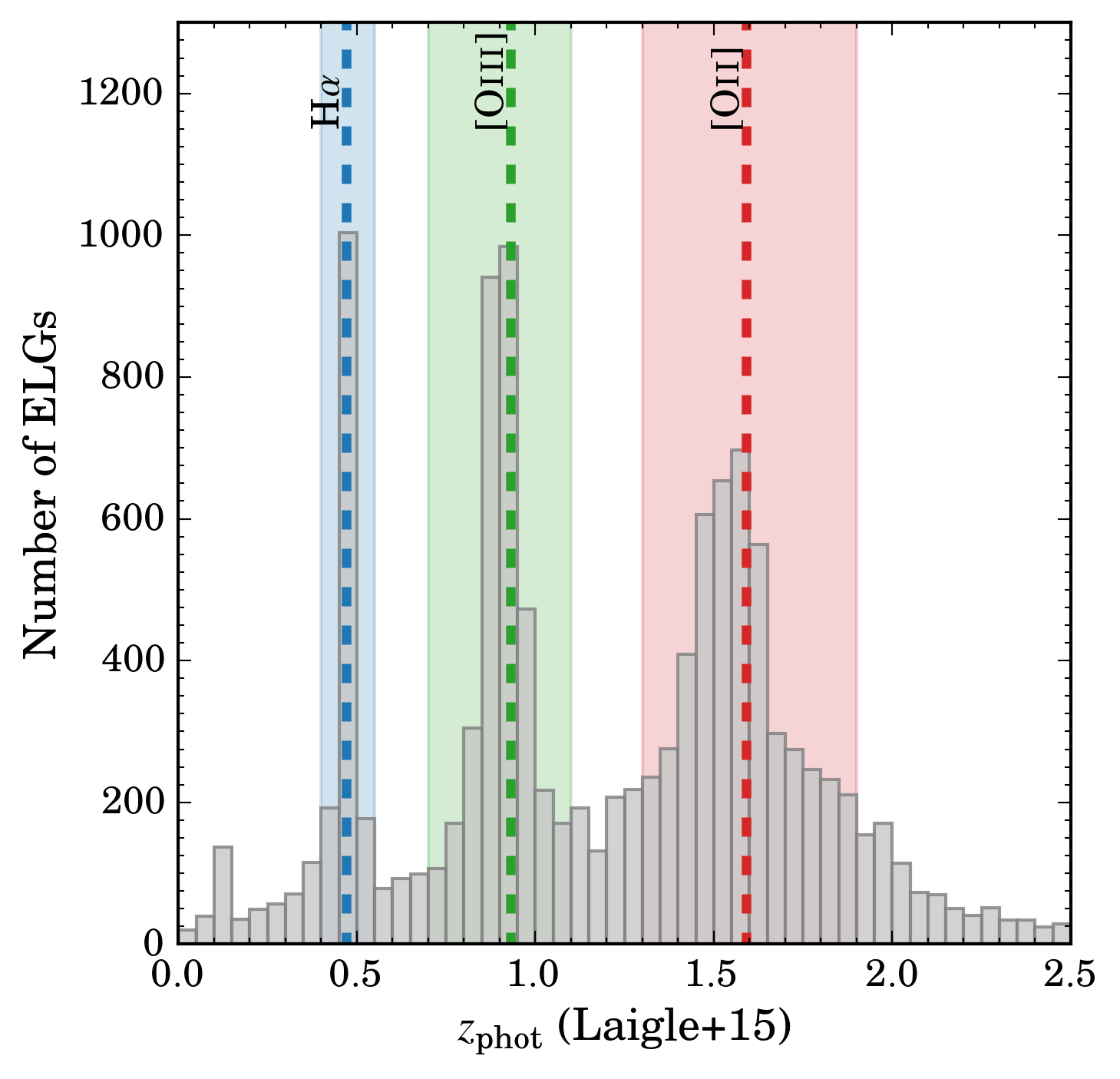}
	\caption{Photometric redshift distribution of ELG candidates. Redshifts used are from \citet{Laigle2016}, which uses a total of 32 photometric filters from the rest-frame UV to near-infrared. Peaks in the redshift distribution correspond to expected redshifts of galaxies with \ha, \oiii, and \oii~emission detection in NB964. The highlighted regions correspond to our photo-$z$ selection range.}
	\label{fig:photoz}
\end{figure}

\subsubsection{Photometric Redshift Selection}

Given the large quantity of multiwavelength observations that have been obtained over the past two decades, reliable and accurate photometric redshifts are now available for use by the community. We use the recent $zYJHK$-selected COSMOS2015 photo-$z$ catalog \citep{Laigle2016}, which uses a combination of 18 broadbands, 12 medium bands, and 2 narrowbands ranging from the rest-frame UV to near-infrared to measure photometric redshifts with $\sigma_{\Delta z/(1+z_\textrm{spec})} \sim 0.007$ in precision. Of the 18,268 ELG candidates, a total of 12,264 of them have measured photometric redshifts by \citet{Laigle2016} ranging from $0 < z < 6$.

Figure \ref{fig:photoz} shows the redshift distribution of our ELG candidates. The pronounced peaks in the distribution correspond to expected redshifts of \ha, \oiii, and \oii~emission line galaxies. We restrict our photo-$z$ selection to the highlighted ranges shown in Figure \ref{fig:photoz} and described in detail in Table \ref{table:criteria}. In total, we select 1150 \ha, 3234 \oiii, and 4625 \oii~emitters based on our photo-$z$ selection. 

\begin{figure}
	\centering
	\includegraphics[width=\columnwidth]{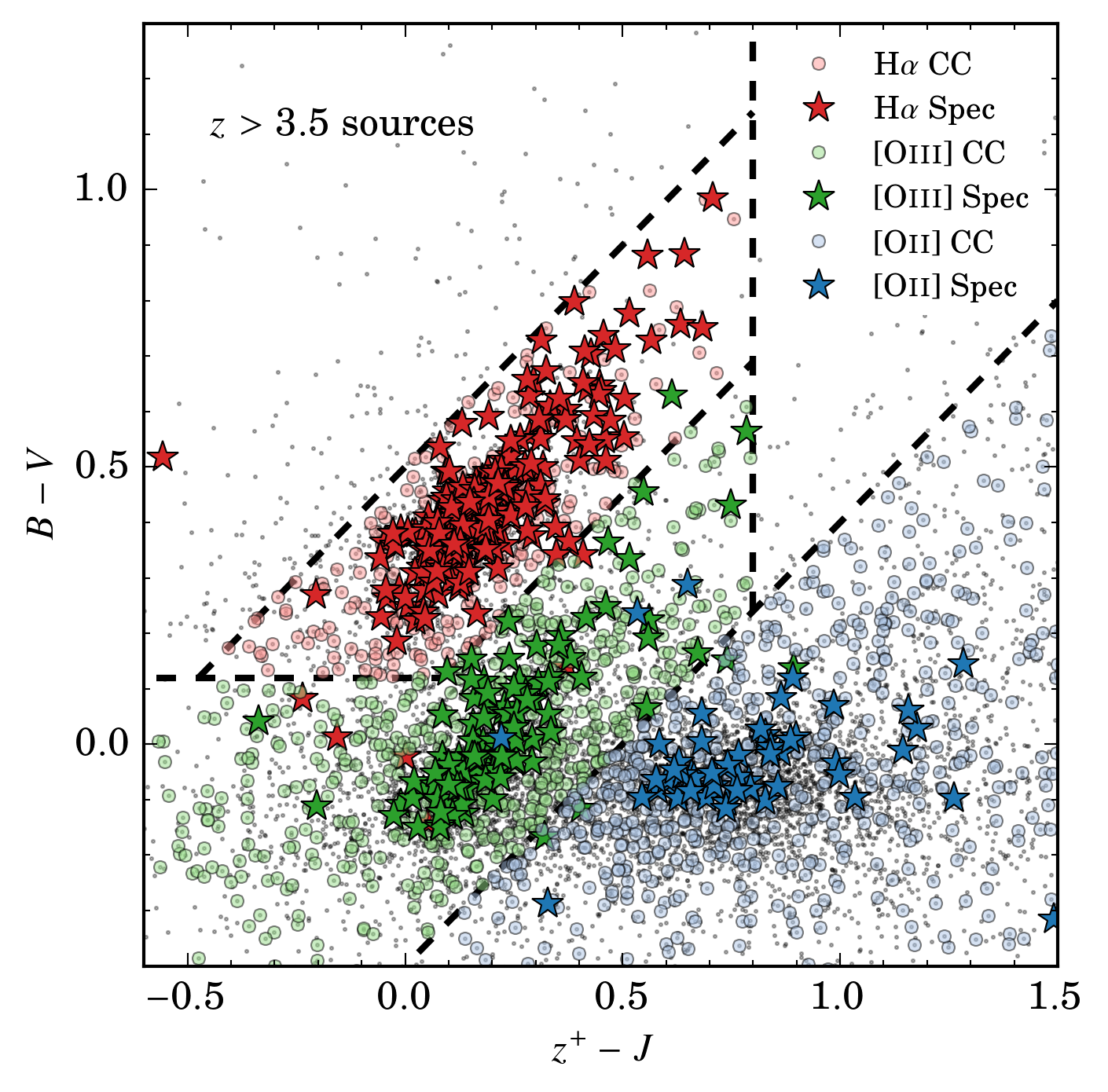}
	\caption{Our $BVzJ$ color-color selection using Subaru/SuprimeCam and VISTA/VIRCAM measurements from the COSMOS2015 catalog \citep{Laigle2016}. We select our color-color diagnostic by identifying the combination which cleanly separates our emission line galaxy samples based on the sources that are spectroscopically confirmed. We identify a region of red $(B-V)$ colors that is linked to $z \sim 3 - 4$ $B$-dropouts. Based on our empirically designed diagnostic, we measure a 6.7\%, 13.2\%, and 19.1\% contamination rate for our \ha, \oiii, and \oii~emitters. The higher contamination in our \oii~samples is due to the difficulty of separating sources that are true \oii~emitters and those that have strong 4000\AA~breaks, which have similar $BVzJ$ colors. Our color-color selection (and/or variations of it) will be used in the other LAGER fields.}
	\label{fig:CC_calib}
\end{figure}

\subsubsection{Color-Color Calibration and Selection}

Although using photometric redshifts selects a few thousand emission line galaxies, it also causes a limitation of our samples to those that have well constrained stellar continuum in a dynamic range of wavelengths. This biases our selection against galaxies with strong nebular emission lines and faint continuum (e.g., low-mass, high equivalent width emitters) which would either have poorly constrained or no photo-$z$ measurements. 

To take this subset of the ELG population into account, we use the distribution of spectroscopically confirmed ELGs to design a color-color calibration that would select sources missed by our photo$-z$ selection. We note that for such a calibration to work requires we maximize the number of sources selected while also minimizing the number of photometric broadbands used.  

We show our $BVzJ$ color-color calibration in Figure \ref{fig:CC_calib} and highlight the selection criteria in Table \ref{table:criteria}. The photometry used is directly from the COSMOS2015 catalog and is based on Subaru/SuprimeCam $BVz$ \citep{Taniguchi2007,Taniguchi2015} and VISTA/VIRCAM $J$ \citep{McCracken2012} imaging. A clear distinction between the spectroscopically confirmed \ha, \oiii, and \oii~emitters is seen in the color-color space. A red cut in $B-V$ colors was placed for \ha~to limit contamination due to potential $z \sim 3 - 4$ $B$-dropouts that may have entered our samples. Of the 12226 COSMOS2015-matched LAGER sources, a total of 8856 sources did not satisfy our spec-$z$ and photo-$z$ criteria. Of these 8856 sources, only 169 ELG candidates lack either $B$, $V$, or $z$ photometry while 1130 sources are missing $J$ photometry. This is due to the wider area covered by the Subaru images in comparison to the UltraVISTA $J$ imaging. In total, we select 204 \ha, 578 \oiii, and 683 \oii~emitters based on their color-colors alone.

\begin{table*}
	\centering
	\caption{Properties of each emission line-selected sample. For each sample, we select emitter candidates in order of spectroscopic redshifts, photometric redshifts, and color-colors. We show the total number of emitters selected per each criteria and the total number of emitters that enters our final sample. Contamination rates are measured by using potential ELGs with spectroscopic redshifts and testing to see how many of them are confirmed \ha, \oiii, or \oii~emitters versus how many are incorrectly identified. Although the color-color selection increases the contamination rate by 1 -- 4 percent, we make up for this with the $\sim 15$ percent increase in the sample size.}
	\begin{tabular*}{\textwidth}{@{\extracolsep{\fill}}cccccccccc}
		\hline
				& 	& \multicolumn{4}{c}{Number of Selected Sources} & \multicolumn{3}{c}{Contamination}\\
		\cline{3-6} \cline{7-9}
		Line	& $z$ & Spec-$z$ & Photo-$z$ & Color-Color & Total & Photo-$z$ & Color-Color & All\\
		\hline
		\ha 	& 	$0.47\pm0.01$ 	& 	222	&	1151	& 	204	&	1577 & 6.2\% & 6.7\% & 6.3\%\\
		\oiii 	&   $0.93\pm0.01$ 	& 	124 & 	3231 	& 	578 &	3933 & 9.0\% & 13.2\% & 9.6\%\\
		\oii 	& 	$1.59\pm0.01$	& 	 59 & 	4625 	& 	683 & 	5367 & 18.6\% & 19.1\% & 18.7\%\\
		\hline
	\end{tabular*}
	\label{table:sample_size}
\end{table*}

\subsection{Contamination}

\subsubsection{Misidentified Lines}
As with any narrowband survey, misidentification of the emission line observed in the narrowband filter does occur. Typical contaminants tend to be galaxies that emit a different emission line than the one mistakenly identified. Another possibility are galaxies where a strong continuum feature (e.g., strong 4000\AA~break) is observed in the NB filter resulting in a color excess imitating a potential emission line galaxy.

To understand the sources of our contaminants and their effect on our sample, we use the wealth of spectroscopic data introduced in \S \ref{sec:spec_sel}. Contamination rates for the photo-$z$ and color-color selections are measured by comparing the number of correctly identified sources to those incorrectly identified via spectroscopic redshifts. The quality flags for the correctly identified emitters are ignored as the narrowband color excess is added confirmation that an emission line is present at the specific spectroscopic redshift (even if the line used to measure the spectroscopic redshift is not the same line observed in NB964; e.g., 3D-{\it HST} G141 \ha~detection at $z = 1.59$ corresponding to \oii~in NB964). We can not argue the same point for those outside of the narrowband filter coverage and therefore require a quality flag $ > 2$ (4 quality flag system) for the `incorrectly' identified sources. 

A total of 17 contaminants were found in our photo$-z$ and CC selection for our \ha~sample. We identify several of them based on their spectroscopic redshifts as 1 \oiii5007, 3 \nii6548, and 2 \nii6583 emitters. Our \oiii~selection resulted in 24 contaminants with 3 \oii~emitters, 4 \ha~emitters, and 1 [Mg{\sc ii}] emitter. We identify 26 contaminants in our \oii~selection with 3 [Ne{\sc iii}], 1 \hb, and 4 \oiii~emitters. One main contribution to the total amount of contamination in our samples are galaxies that exhibit a false emission line. This primarily occurs with our \oii~sample where the narrowband filter may observe the 4000\AA~break while the majority of the broadband filter covers the continuum bluewards of the break. This results into a source imitating an emission line galaxy solely due to color. We identify only 2 sources which have redshifts consistent with the 4000\AA~break within the NB964 wavelength coverage. 

Overall, we estimate the contamination rates for our samples as 6.3\%, 9.6\%, and 18.7\% for our \ha, \oiii, and \oii~samples, respectively, as shown in Table \ref{table:sample_size}. The contamination rates based on our photo-$z$ selection alone is about $1 - 4$ percent lower than the color-color selection, although the latter increases the sample sizes by $\sim 15$ percent per each emission line. In this sense, we are significantly increasing our sample size and also taking into account sources with unreliable or no photo-$z$ measurements at the cost of a marginal increase in contamination.

\subsubsection{Active Galactic Nuclei}
\label{sec:AGNs}
Although strong nebular emission lines are powered by the ionizing power of stars formed within H{\sc ii} regions, they can also be produced by the hot, ionizing radiation emitted by the accretion processes of active galactic nuclei (AGN). In order for us to investigate the statistical star-forming properties of our samples, we need to understand the level of contribution arising from AGN activity.

To assess the level of AGN contamination, we first match our samples to identified X-ray sources in the Laigle catalog which are drawn from {\it XMM}-COSMOS \citep{Cappelluti2007,Hasinger2007,Brusa2010} and {\it Chandra}-COSMOS \citep{Elvis2009,Civano2012}. We find a total of 3, 10, and 32 X-ray detections for our \ha, \oiii, and \oii~samples, respectively, which would suggest that X-ray AGNs contribute $< 1\%$ to our samples. 

\begin{figure}
	\centering
	\includegraphics[width=\columnwidth]{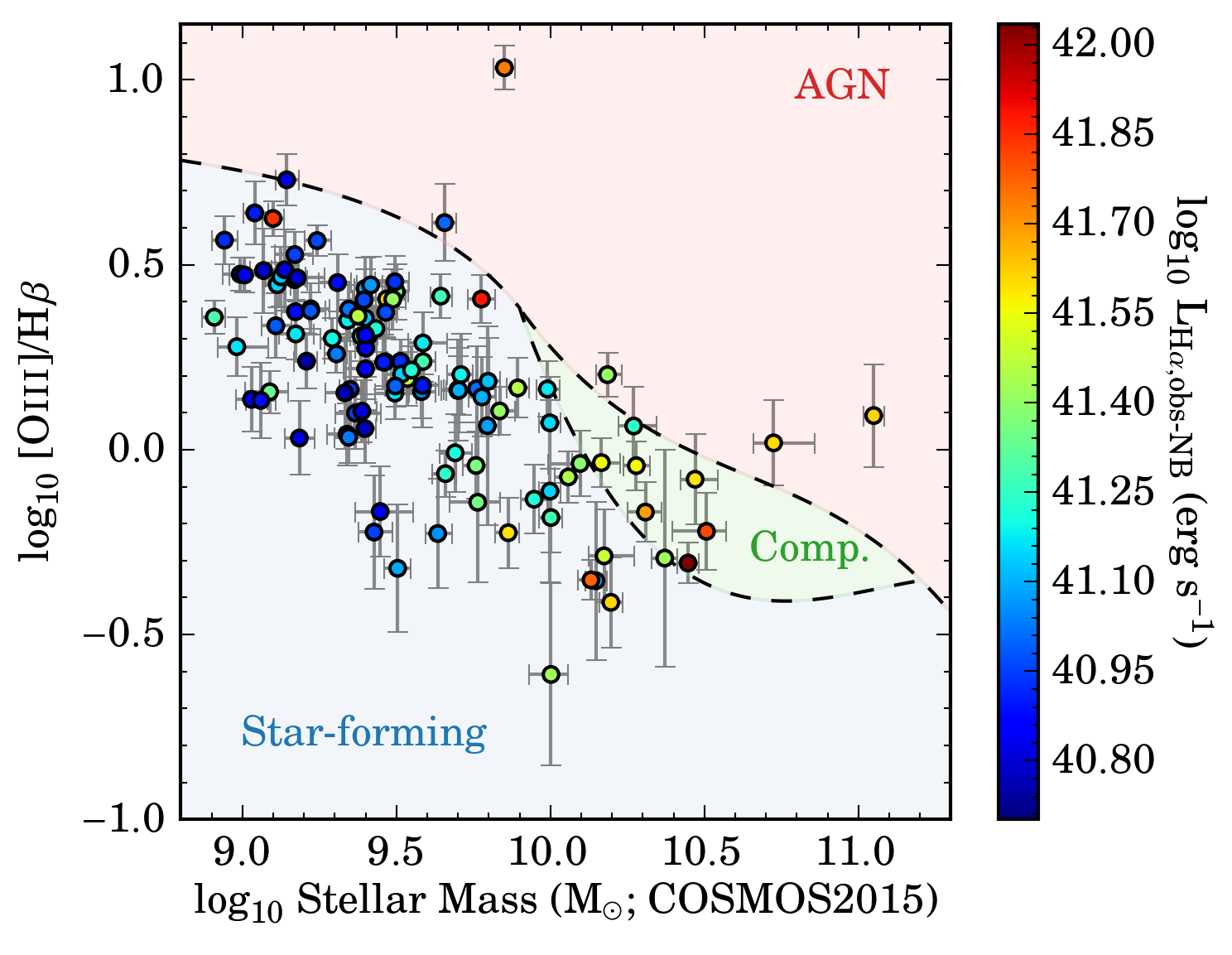}
	\caption{The Mass-Excitation(MEx) diagram of our $z = 0.47$ \ha~emitters with available $z$COSMOS spectra. The \citet{Juneau2011} selection regions separate the samples into star-forming, AGN, and composite (star-forming+AGN) populations. We find that the majority of our \ha~emitters are star-forming galaxies, with 6 \ha~emitters classified as AGNs and 9 as composites. Shown in the colorbar is the observed \ha~luminosity measured from the narrowband photometry. We find that the \ha~emitters classified as AGNs and composites are mostly the brighter emitters with $L > 10^{41.55}$ erg s$^{-1}$. This could suggest that bright \ha~populations have higher AGN fractions, in line with recent spectroscopic results (e.g., \citealt{Sobral2015}), although we note the selection function of our $z$COSMOS-matched sources is more complicated.}
	\label{fig:MEx_diagram}
\end{figure}

We also use a mid-infrared color diagnostic that utilizes the 1.6\micron~bump of star-forming galaxies and the power-law shape of AGN spectral energy distributions (SEDs) to assess AGN contamination. This is a similar approach used in infrared AGN selection (e.g., \citealt{Stern2005,Donley2012}) and has been used in several narrowband studies (e.g., \citealt{Garn2010,Sobral2013,Sobral2015,Khostovan2015}). First, we assume a minimum spectral slope of $\beta = -0.5$ in the AGN SEDs ($f_\nu \propto \nu^\beta$; e.g., \citealt{Alonso2006,Donley2007,Donley2012}). We then select two adjacent filters that are redwards of the 1.6\micron~bump, which arises from the stellar atmospheres of cool stars in star-forming galaxies \citep{Sawicki2002}. Red colors would signify a rising SED representative of an AGN, while blue colors represent star-forming galaxies. Using CFHT/WIRcam $K_s$ \citep{McCracken2010} and SPLASH {\it Spitzer} IRAC photometry (Capak et al., in prep), we set the limits as such: \ha~[$K_s - 3.6$] > 0.27, \oiii~[$3.6 - 4.5$] > 0.13, \oii~[$4.5 - 5.6$] > 0.13 mag. For the \ha~sample, we use a filter slightly bluewards of the 1.6\micron~bump ($K_{s}$ traces rest-frame 1.3 - 1.6 \micron~while IRAC CH1 traces rest-frame 2.1 -- 2.6 \micron). This is due to the large wavelength gap between ground-based imaging and {\it Spitzer} coverage. Our AGN contamination rates are 10\%, 13\%, and 23\%, for \ha, \oiii, and \oii, respectively.

For our \ha~sample, we also investigate the AGN fraction using nebular diagnostics and the $z$COSMOS spectra \citep{Lilly2009}. Unfortunately, the \ha~line at $z = 0.47$ is close to the red boundary of the VISTA/VIMOS coverage such that we do not have \ha~and \nii~coverage for all our spectroscopically confirmed samples and, therefore, can not use the traditional BPT diagram \citep{Baldwin1981}. Given this limitation, we investigate our AGN fractions using the Mass-Excitation (MEx) diagnostic \citep{Juneau2011} which depends on the \oiii/\hb~ratio (from the $z$COSMOS spectra) and stellar mass (from the COSMOS2015 catalog). We restrict our analysis to only those sources for which have photometric redshifts $\pm 0.05$ from the spectroscopic redshift to ensure that the stellar mass used is consistent with the corresponding spec-$z$. Figure \ref{fig:MEx_diagram} shows the MEx diagram along with 118 H$\alpha$ emitters with $z$COSMOS spectra and is color-coded with the NB-observed H$\alpha$ luminosity. We find 6/118 \ha~emitters classified as AGN corresponding to $\sim 5\%$ and 9/118 are classified as composite AGN-SF. If we consider the combination of the two as AGN contaminants, then our AGN fraction is $\sim 13\%$ similar to what we measure based on our IR selection. Overall, we can assume that our \ha~AGN fraction rests somewhere around 5 - 10\% of our sample. Although the MEx diagnostic was calibrated on continuum-selected samples at $z < 0.2$ and our samples are ELG-selected, the sources shown in Figure \ref{fig:MEx_diagram} are $z$COSMOS-spectroscopically confirmed ELGs, which is an $i-$band continuum-selected study.

We note that our measurements presented here represent our assessment for the full sample, while it has been reported that the AGN fractions are distributed such that the brightest emission line galaxies typically are AGN dominated (e.g., \citealt{Sobral2015}). Figure \ref{fig:MEx_diagram} also suggests that the bright \ha~emitters are primarily AGNs/composites, although the $z$COSMOS-LAGER matched selection function is more complicated. Spectroscopic follow-up of our samples is needed to provide us with a better gauge on the AGN fractions and how they are distributed throughout our samples.

\subsection{Are we selecting \oiii~or \hb?}
\label{sec:contrib_hbeta}

The observer frame separation between the two \oiii~lines and \hb~is about 185\AA~and 280\AA~for \oiii4959 and \oiii5007, respectively, which is significantly larger than the 95\AA~FWHM of the NB964 filter. Although this suggests our \oiii~line fluxes have no blending issues with the \hb~line, it does raise the issue of having \hb~emitters populating our samples and misidentified as \oiii~emitters. This is in part due to the selection technique used where photo-$z$ and color-color selection cover too wide of a wavelength range to decouple the two lines. 

Previous \oiii~narrowband surveys find that \hb~emitters tend to contribute more towards fainter line luminosities depending on survey properties (e.g., $10 - 20\%$; \citealt{Sobral2015,Khostovan2016}). Using the spectroscopic measurements from the literature, we find that from the 124 spectroscopically confirmed sources in our \oiii~sample, only 4 are found to be \hb~emitters. We note that spectroscopic surveys have different selection functions in respect to our survey, such that they are biased to the brightest line fluxes. Since \hb~emitters are found to be misidentified as faint \oiii~emitters, we stress caution when interpreting the number statistics solely on the lack of spectroscopic coverage of our fainter \oiii-selected emitters which could actually be \hb~emitters. As we show later in \S\ref{sec:O3_LF}, the contribution of \hb~emitters is negligible and only affects our measured luminosity functions in the faintest luminosity bins where the number densities slightly decrease but are still within $1\sigma$ error.

\section{Methodology}
\label{sec:methodology}

\subsection{\nii~Contamination of \ha~Sample}

A fraction of narrowband flux measured in our \ha~samples is contaminated by the nearby \nii6583 line. Figure \ref{fig:NB_profile} shows an example using $z$COSMOS 1D spectra of one of our $z = 0.47$ \ha-selected emitters overlaid with the NB964 filter profile. Both \nii6548,6583~lines are observed within the filter, such that our \ha~samples are susceptible to \nii~contamination. Not correcting for this contamination would result into overestimated line luminosities that systematically shift our luminosity functions.

Several studies have corrected for this contamination by assuming a fixed \nii/\ha~line ratio (e.g., \citealt{Ly2007,Morioka2008}) based on spectroscopic observations of local, nearby star-forming galaxies (e.g., \citealt{Kennicutt1992,Gallego1997}). Observations of $z < 2.5$ star-forming galaxies have reported a wide range in \nii/\ha~line ratios ranging between 0.02 -- 0.4 (e.g., \citealt{Steidel2014}), such that a constant line ratio may not best represent our samples. Recent works have modeled the redshift evolution of the \nii/\ha~line ratio empirically as dependent on stellar mass \citep{Faisst2018} and through simulations \citep{Merson2018}. We instead apply the \nii/\ha~calibration designed by \citet{Villar2008} and recomputed by \citet{Sobral2012}: $\log_{10} \textrm{\nii}/\textrm{\ha} = -0.924 + 4.802x -8.892x^2 + 6.701x^3 - 2.27x^4 + 0.279x^5$, where $x = \log_{10} \textrm{EW}_\textrm{rest}(\textrm{\ha}+\textrm{\nii})$, which is directly measured from our NB observations.

Although calibrated at $z\sim 0$, this equivalent width-based correction is shown to be consistent with the \nii/\ha~line ratios of NB-selected \ha~emitters up to $z \sim 1.5$ (e.g., \citealt{Sobral2015}). The \nii/\ha~correction applied to our \ha~sample results in a median reduction of the observed line fluxes by $0.11\pm0.02$ dex.

\subsection{Completeness Correction}
\label{sec:completeness}
Knowing the completeness limits of our samples as a function of line flux is imperative for calculations of the luminosity functions. Towards lower line fluxes, our observations and sample selections will miss an increasing number of emitters. Not correcting for this effect results in an underestimation of the number densities of emission line galaxies towards the faint-end of the luminosity function.

We measure our completeness limits using an empirical approach similar to that used in various \ha, \oiii, and \oii~studies (e.g., \citealt{Sobral2013,Sobral2015,Khostovan2015}) by running simulations of emission line galaxies covering a wide range of line flux and taking into account selection effects. This is done by using our sample of $\sim 2.5$ million galaxies not selected as ELGs based on our nebular color excess cuts ($\Sigma < 3$, EW$_\textrm{obs} < 52.5$\AA, and NB $ > 25.45$ mag) as our mock sample. The advantage of this approach is that our measurements are empirical-based and therefore independent of modeling artificial sources and their nebular and stellar properties, especially when it comes to measuring the recovery fractions based on the redshift and color-color selection criteria. 

\begin{table}
	\caption{The 5$\sigma$ and $30\%$ completeness line luminosity limits of our survey. The 5$\sigma$ narrowband magnitude limit is 25.45 mag, corresponding to a limiting line flux of $8.2\times 10^{-18}$ erg s$^{-1}$ cm$^{-2}$. Our results shown in this paper are based on emission line galaxies above the $30\%$ line luminosity limit.}
	\begin{tabular*}{\columnwidth}{@{\extracolsep{\fill}}ccc}
		\hline
		Line & $\log_{10} L_{lim,5\sigma}$ & $\log_{10} L_{30\%,comp}$\\
		&  (erg s$^{-1}$) & (erg s$^{-1}$)\\
		\hline
		\ha & 39.83 & 40.14\\
		\oiii & 40.55 & 40.85\\
		\oii & 41.13 & 41.27\\
		\hline
	\end{tabular*}
	\label{table:comp_limits}
\end{table}

\begin{figure}
	\centering
	\includegraphics[width=\columnwidth]{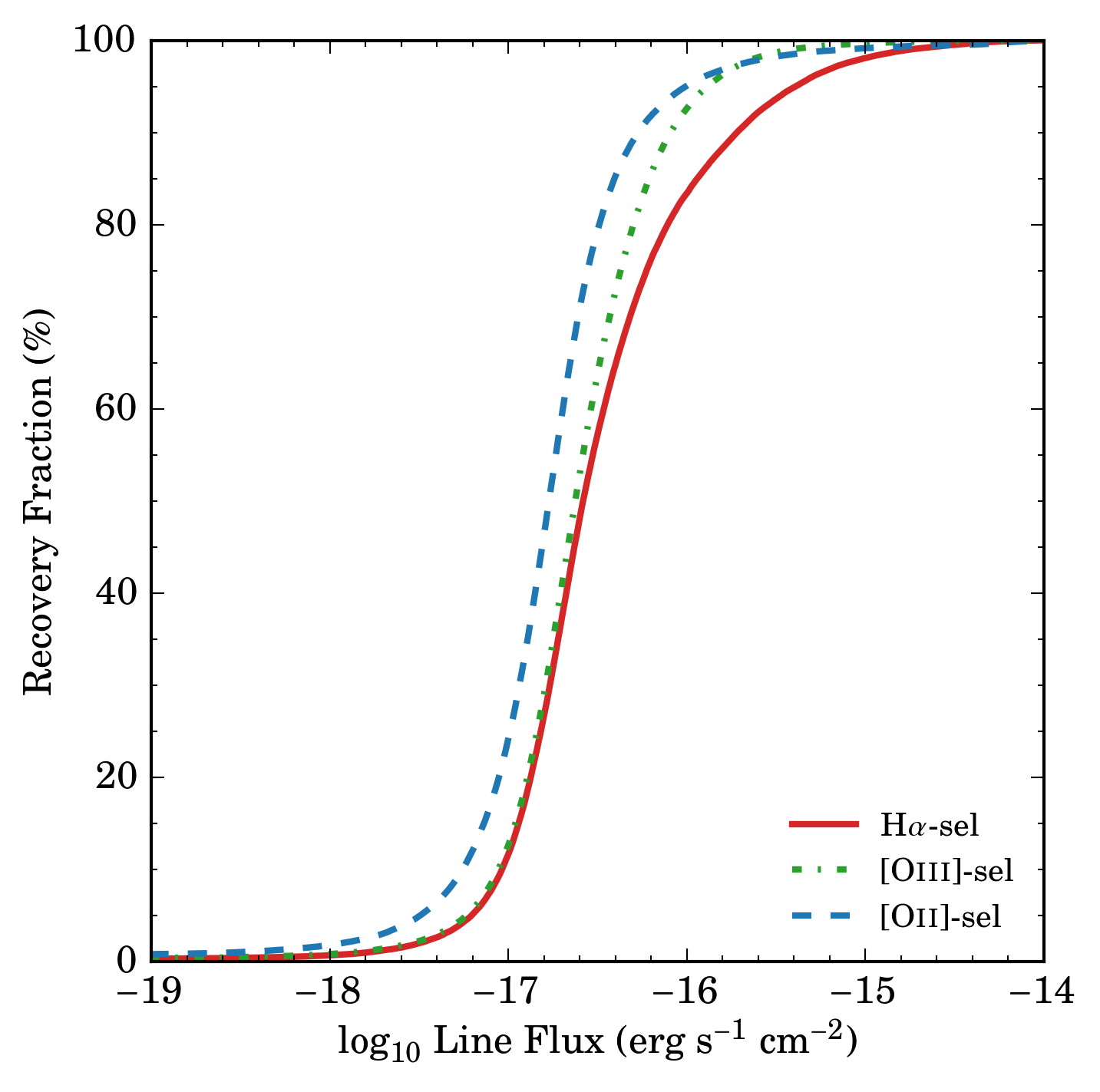}
	\caption{The completeness corrections for all three emission line detections as dependent on observed line flux. The 30 percent completeness limits are $-16.77$, $-16.79$, and $-16.94$ dex in line flux, which corresponds to $10^{40.14}$, $10^{40.85}$, and $10^{41.27}$ erg s$^{-1}$ in observed line luminosity for \ha, \oiii, and \oii-selected sources, respectively.}
	\label{fig:completeness}
\end{figure}

We use our sample of $\sim 673000$ COSMOS2015-matched non-ELGs as our mock sample to measure the necessary completeness corrections. Each of the non-ELGs have multi-wavelength photometry and photometric redshifts from the COSMOS2015 catalog and a subset of them have spectroscopic redshifts from the compilation described in \S\ref{sec:spec_sel}. Using our line identification criteria outlined in \S\ref{sec:sample_selection}, we subdivide the non-ELG sample into \ha, \oiii, and \oii~subsamples. These are sources that have the potential of being an ELG of the respective line (spectroscopic redshift, photometric redshift, or color-color selected) but failed to pass our narrowband color excess criteria highlighted in \S\ref{sec:ELG_selection}. 

To measure the completeness of each respective emission line sample, we apply a fake line to each mock galaxy by convolving the line flux with the narrowband and broadband filter profiles, which are then used to measure the observed narrowband and broadband magnitudes. Second, we use the mock sample that now has a fake emission line added to each source and apply our ELG selection criteria, as defined in \S\ref{sec:ELG_selection}. This then gives us the total number of recovered \ha, \oiii, and \oii~emitters at a given line flux. We repeat these steps by increasing the inputted fake line flux from $10^{-20}$ to $10^{-14}$ erg s$^{-1}$ cm$^{-2}$ in increments of 0.01 dex. The completeness of each emission line sample is defined as the total number of recovered sources at any given line flux divided by the total number of sources recovered at $10^{-14}$ erg s$^{-1}$ cm$^{-2}$, where the number of sources asymptotes. Using this empirical approach, we are assessing for a given mock line flux (and also the stellar continuum of each source), how many sources do we potentially miss based on not only the ELG candidate selection, but also when identifying candidates based on \ha, \oiii, and \oii.

Figure \ref{fig:completeness} shows the recovery fraction of our samples in terms of observed line flux. We find 30\% completeness limits of $10^{-16.77}$, $10^{-16.79}$, and $10^{-16.94}$ erg s$^{-1}$ cm$^{-2}$ for our \ha, \oiii, and \oii~samples, respectively, as shown in Table \ref{table:comp_limits}. We incorporate the curves in our measurements of the luminosity functions as described in \S \ref{sec:make_lfs} and include 20\% of the total completeness corrections in quadrature to the luminosity functions.

\subsection{Filter Profile \& Volume Corrections}
\label{sec:filter_volume_corr}
As shown in Figure \ref{fig:NB_profile}, the narrowband filter profile is not a top-hat, which has implications on the observed luminosity distribution. The `intrinsically' faint (luminous) line emitters would be detectable only within a narrow (wide) range around the center of the filter profile curve. This means that the faint (luminous) emitters will cover smaller (larger) volumes in respect to that measured using a simple top-hat. Furthermore, the `intrinsically' luminous emitters can be detected further from the center of the filter profile and would be observed as faint emitters due to the low transmission. The effect this has on the shape of the luminosity function is that the faint-end (bright-end) will have overestimated (underestimated) number densities.

To correct for this effect, we use our completeness-corrected luminosity functions and make an initial Schechter fit. We then generate $10^7$ mock sources by randomly assigning luminosities based on the initial luminosity function and redshifts that cover the full range of the narrowband profile. The line luminosities of these sources are then convolved with the narrowband filter and a top-hat filter with width equivalent to the FWHM of the narrowband. The top-hat filter would fully recover the inputted line luminosity within the redshift range, while the narrowband filter would have a reduction (inflation) of luminous (faint) mock sources due to their position along the filter profile. The ratio of the mocks sources recovered by the top-hat and narrowband filters per luminosity bin defines the correction factor needed to take into account the filter profile effects. These corrections increase from faint to bright line luminosities and can be as high as 35\%, 50\%, and 65\% increase in the number densities at the brightest line luminosities for \ha, \oiii, and \oii, respectively.

\subsection{Dust Correction}
\label{sec:dust_calib}

Measuring the intrinsic line luminosities is ideal when investigating the statistical properties of emission line galaxies. Unfortunately, dust absorption along the line-of-sight causes a decrease in the line luminosities when observing a source. Although this effect can be corrected, numerous factors can affect the required corrections, such as the metal enrichment history of the galaxy, its star formation activity, and dust geometry along the line of sight. In emission line galaxy studies, a constant $A_\textrm{\ha} = 1$ mag dust extinction is typically assumed (e.g., \citealt{Hopkins2004,Takahashi2007,Sobral2013,Matthee2017}) and has been shown to be reliable up to $z \sim 2$ (e.g., \citealt{Sobral2012,Ibar2013,Momcheva2013,Sobral2016b}). Empirical dust calibrations have also been designed in terms of observed star formation rates (e.g., \citealt{Hopkins2001}), rest-frame colors (e.g., \citealt{Sobral2012}), and stellar mass (e.g., \citealt{Garn2010}). In this paper, we assume two different cases for dust corrections: 1) a constant $A_\textrm{\ha} = 1$ mag extinction and 2) our own empirically designed calibration based on rest-frame SDSS $g - r$ color. We assume a \citet{Calzetti2000} dust attenuation curve for both cases. In the case of the empirical calibration, we use archival $z$COSMOS 1D spectra and SDSS DR12 measurements to design and test our calibration.

\begin{figure}
	\centering
	\includegraphics[width=\columnwidth]{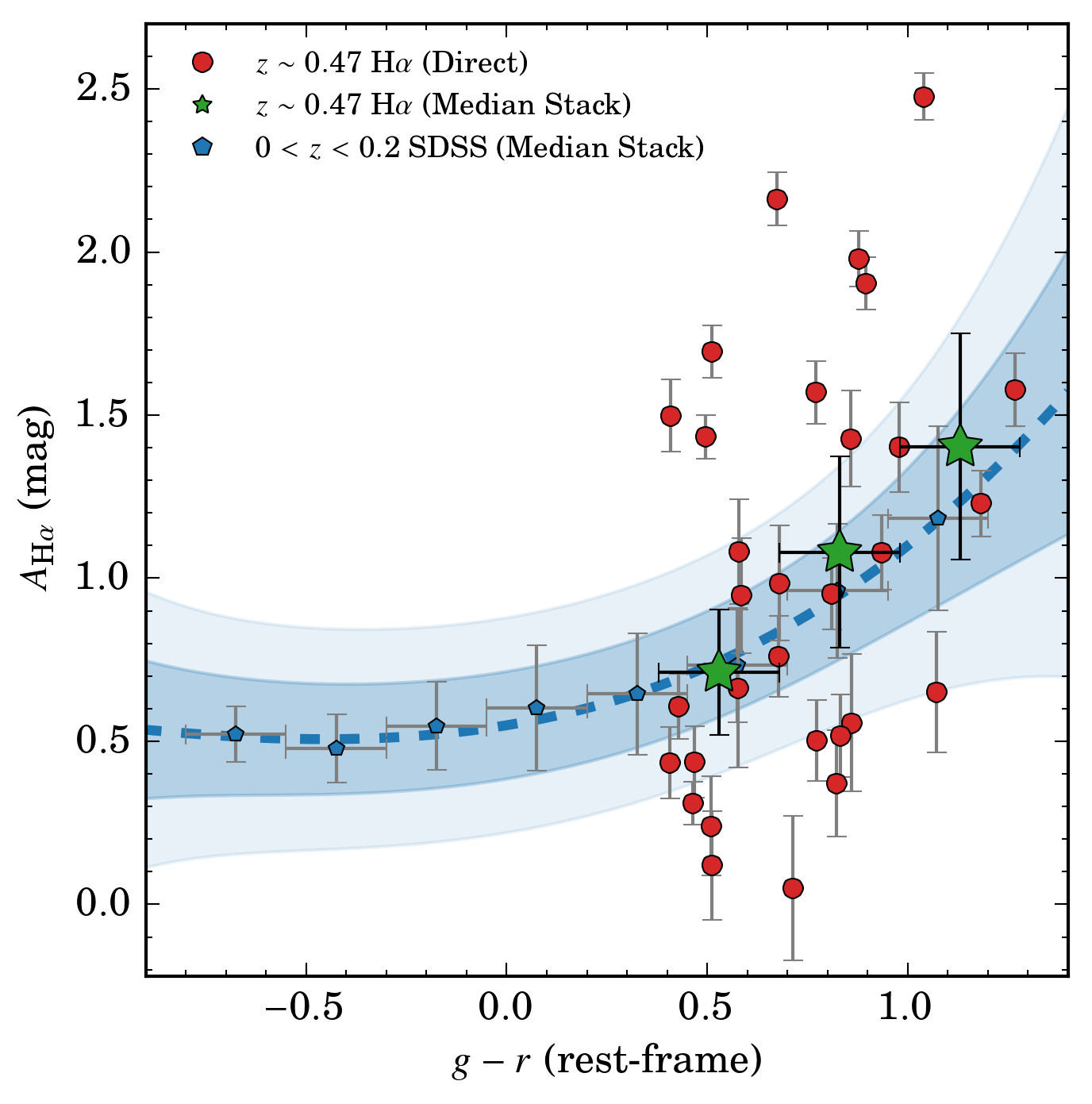}
	\caption{Our empirically designed dust calibration. We show the median \aha~ for local $z \sim 0.1$ \ha~emitters in the SDSS DR12 survey \citep{Thomas2013} as dependent on their rest-frame $(g - r)$ colors as shown in {\it blue pentagons}. A third-order polynomial is fitted to the median stacks in order to design an empirical dust calibration with the {\it dark} and {\it light} shaded regions representing the 1$\sigma$ and 2$\sigma$ regions, respectively. We test our calibration against our $z = 0.47$ \ha~emitters with $z$COSMOS spectra, shown as {\it red circles}, and find an agreement when comparing the median \aha~of our samples ({\it green stars}) rather than comparing individual \ha~emitters to the calibration. This would suggest that such a calibration could be used to correct our \ha~emitters when measuring the luminosity functions.}
	\label{fig:dust_calib}
\end{figure}

We use the publicly available SDSS DR12 spectroscopy measured by the Portsmouth group and select only those classified as star-forming based on the BPT diagnostic \citep{Thomas2013}. \ha~sources are selected by setting a rest-frame equivalent width cut corresponding to our survey and within a redshift range of $0 < z < 0.2$, corresponding to a 2 Gyr timeframe. The \ha~extinction is measured using Balmer decrements:
\begin{eqnarray}
	\centering
	A_\textrm{\ha} = 2.5\frac{k(\textrm{\ha})}{k(\textrm{\hb}) - k(\textrm{\ha})} \log_{10}\frac{F_\textrm{\ha}/F_\textrm{\hb}}{2.86}
	\label{eqn:Balmer}
\end{eqnarray}
where \aha~is the \ha~extinction, $F_\textrm{\ha}$ and $F_\textrm{\hb}$ are the \ha~and \hb~observed fluxes, $k(\textrm{\ha})$ and $k(\textrm{\hb})$ are the reddening curves from the \citet{Calzetti2000} dust attenuation curve for \ha~and \hb, respectively. An intrinsic \ha/\hb~ratio of 2.86 is assumed and corresponds to a temperature of $T = 10,000$K and an electron density $n_e = 100$ cm$^{-3}$ for case B recombination \citep{Osterbrock1989}. 

To limit the effects of uncertain \aha~measurements in the SDSS data, we limit the samples such that $\textrm{\aha}/\sigma_\textrm{\aha} > 2$. As we are interested in the typical dust properties of galaxy samples rather than individual galaxies, we bin our SDSS calibration sample in $g - r$ color and apply a third-order polynomial fit with the best-fit being:
\begin{eqnarray}
\centering
A_\textrm{\ha} = 0.026 x^3 + 0.295 x^2 + 0.249 x + 0.557
\end{eqnarray}
with $x = (g-r)$ rest-frame color being a proxy for stellar continuum (e.g., \citealt{Groves2012}). Our empirical dust-correction prescription is calibrated within the range of $-0.8 < (g - r) < 1.3$ mag.

To test our $z \sim 0.1$ dust calibration at higher redshifts, we use the publicly available, flux-calibrated 1D spectra from the $z$COSMOS survey of our $z = 0.47$ \ha~samples \citep{Lilly2009}. Line fluxes are measured by fitting Gaussian profiles centered on the \ha~and \hb~emission lines and then integrating the profiles. The \ha~extinction is then measured using Equation \ref{eqn:Balmer}. Of 125 \ha~emitters with $z$COSMOS spectra, we find only 31 of them had reliable \ha~and \hb~measurements. This is primarily due the \ha~line at $z = 0.47$ falling very close to the red edge of the VLT/VIMOS wavelength range. 

Figure \ref{fig:dust_calib} shows our SDSS dust calibration along with the individual $A_\textrm{\ha}$ measurements and median stacks. We find a wide scatter in the individual $A_\textrm{\ha}$ measurements which suggests our samples have a diverse range of dust properties, although the selection function here is more complex. The median stacks of the individual detections show consistent $A_\textrm{\ha}$ in comparison to our SDSS calibration, suggesting that our approach can be applied for our samples as a whole.

A total of 1543 out of 1577 \ha~emitters have Subaru $R$ (rest-frame $g$) and $z^{++}$ (rest-frame $r$) detections for which 24 \ha~emitters are outside of our calibration range and 19/1577 \ha~emitters have either $R$ or $z^{+}$ detection. For these 43 sources, we use the limits of our calibration to measure \aha~(e.g., sources with $g - r$ > 1.3 have \aha~measured at $g - r = 1.3$ mag). Applying our calibration, we find that the typical dust correction for the $z \sim 0.47$ \ha~sample is \aha$=0.76\pm0.14$ mag, which is somewhat lower than the $1$ mag extinction typically assumed in the literature. We apply both methods of correcting for dust in our luminosity function measurements as described below.

The dust calibrations are also applied to our \oiii~and \oii~samples, but it is unclear whether such a calibration works at their respective redshifts. \citet{Hayashi2013} used $z = 1.47$ dual narrowband survey strategy selecting 809 star-forming galaxies with \ha~and \oii~emission and found that \oii-selected narrowband emitters are biased towards dust-poor systems with typical \aha$ = 0.35$ mag. \citet{Hayashi2015} followed-up 118 $z = 1.47$ \oii~emitters in the Subaru Deep Field with Subaru/FMOS and found typical \aha$ = 0.61$ mag. Using our dust calibrations, we find a median dust extinction correction of \aha$ = 0.55\pm0.12$ mag for our \oii~sample. A comparative study of dust corrections for \oiii~emitters is still lacking. The median \oiii~dust extinction correction using our calibration is \aha$ = 0.60\pm0.10$ mag. Given these limitations, we caution the reader when interpreting the results when the calibration is used for the \oiii~and \oii~samples. 

\begin{table*}
	\centering
	\caption{Best-fit Schechter parameters for each of our samples with the characteristic number density ($\phi^\star$), characteristic luminosity ($L^\star$), and faint-end slope ($\alpha$) treated as free parameters with the luminosity density, $\rho_L$, measured as the full integration of the LF. We measure our luminosity functions based on three different dust prescriptions: observed, constant \aha$=1$ mag, and our own empirical dust calibration. For the case of our \oii~emitters, we set \aha$=0.35$ mag as suggested by \citet{Hayashi2013}. The comoving volumes for each sample are measured assuming a tophat filter with FWHM equivalent to the narrowband, although we take filter profile \& volume corrections into account when measuring the luminosity functions.}
	\begin{tabular*}{\textwidth}{@{\extracolsep{\fill}}c c c c c c c c}
		
		\hline
		Line & $z$ & Comoving Volume & Dust Presc. & $\log_{10} \phi^\star$ & $\log_{10} L^\star$ & $\alpha$  &  $\rho_L$\\
		&    & ($10^5$ Mpc$^{3}$) & & (Mpc$^{-3}$) & (erg s$^{-1}$) & & (erg s$^{-1}$ Mpc$^{-3}$) \\
		\hline
		\ha & $0.47\pm0.01$ & 1.127 &  none & $-3.19^{+0.24}_{-0.30}$ & $41.73^{+0.20}_{-0.16}$ & $-1.77^{+0.12}_{-0.11}$ & $39.14^{+0.18}_{-0.10}$\\
		-- & -- & -- & const & $-3.20^{+0.22}_{-0.26}$ & $42.13^{+0.17}_{-0.15}$ & $-1.78^{+0.11}_{-0.10}$ & $39.55^{+0.17}_{-0.09}$\\
		-- & -- & -- & calib & $-3.36^{+0.25}_{-0.31}$ & $42.21^{+0.22}_{-0.17}$ & $-1.77^{+0.10}_{-0.09}$ & $39.44^{+0.14}_{-0.08}$\\
		\oiii & $0.93\pm0.01$ & 3.434 & none & $-2.13^{+0.20}_{-0.27}$ & $41.36^{+0.15}_{-0.13}$ & $-1.57^{+0.35}_{-0.30}$ & $39.55^{+0.43}_{-0.18}$\\
		-- & -- & --  & const & $-2.22^{+0.25}_{-0.25}$ & $41.95^{+0.14}_{-0.15}$ & $-1.73^{+0.29}_{-0.19}$ & $40.26^{+0.46}_{-0.23}$\\
		-- & -- & -- & calib & $-2.37^{+0.26}_{-0.22}$ & $41.82^{+0.12}_{-0.15}$ & $-1.79^{+0.29}_{-0.15}$ & $40.10^{+0.51}_{-0.28}$\\
		\oii & $1.59\pm0.01$ & 6.732 & none & $-2.08^{+0.14}_{-0.16}$ & $41.73^{+0.09}_{-0.08}$ & $-1.58^{+0.30}_{-0.27}$ & $39.98^{+0.41}_{-0.18}$\\
		-- & -- & -- & const & $-2.02^{+0.11}_{-0.14}$ & $41.95^{+0.08}_{-0.07}$ & $-1.47^{+0.23}_{-0.23}$ & $40.15^{+0.19}_{-0.11}$\\
		-- & -- & -- & calib & $-1.98^{+0.08}_{-0.11}$ & $42.07^{+0.06}_{-0.05}$ & $-1.21^{+0.15}_{-0.20}$ & $40.16^{+0.08}_{-0.04}$\\
		\hline
	\end{tabular*}
	\label{table:LF_free}
\end{table*}

\begin{table*}
	\centering
	\caption{Our best-fit Schechter parameters with fixed $\alpha$ for our observed, constant \aha-corrected, and empirical dust calibration-corrected lumnosity functions. For each we show the characteristic number density, \pstar, and luminosity, \lstar.}
	\begin{tabular*}{\textwidth}{@{\extracolsep{\fill}}c c c c c c c c c}
		
		\hline
		& & \multicolumn{2}{c}{Observed} & \multicolumn{2}{c}{Constant $A_\textrm{\ha}$} & \multicolumn{2}{c}{Dust Calibration} & \\
		\cline{3-4} \cline{5-6} \cline{7-8}
		Line & $z$ & $\log_{10} \phi^\star$ & $\log_{10} L^\star$ & $\log_{10} \phi^\star$ & $\log_{10} L^\star$ & $\log_{10} \phi^\star$ & $\log_{10} L^\star$ & $\alpha$  \\
		&  & (Mpc$^{-3}$) & (erg s$^{-1}$) & (Mpc$^{-3}$) & (erg s$^{-1}$) & (Mpc$^{-3}$) & (erg s$^{-1}$) &  \\
		\hline
		\ha & $0.47\pm0.01$ & $-3.16^{+0.09}_{-0.09}$ & $41.72^{+0.09}_{-0.09}$ & $-3.14^{+0.09}_{-0.09}$ & $42.10^{+0.09}_{-0.09}$ & $-3.32^{+0.10}_{-0.10}$ & $42.19^{+0.11}_{-0.11}$ & $-1.75$ (fixed)\\
		\oiii & $0.93\pm0.01$ & $-2.16^{+0.10}_{-0.12}$ & $41.38^{+0.07}_{-0.06}$ & $-2.12^{+0.10}_{-0.10}$ & $41.90^{+0.07}_{-0.06}$ & $-2.20^{+0.09}_{-0.08}$ & $41.73^{+0.05}_{-0.05}$ & $-1.60$ (fixed)\\
		\oii & $1.59\pm0.01$ & $-1.97^{+0.07}_{-0.07}$ & $41.66^{+0.03}_{-0.03}$ & $-1.95^{+0.06}_{-0.06}$ & $41.90^{+0.03}_{-0.03}$ & $-2.02^{+0.05}_{-0.05}$ & $42.10^{+0.03}_{-0.03}$ & $-1.30$ (fixed)\\
		\hline
	\end{tabular*}
	\label{table:LF_fixed}
\end{table*}

\subsection{Constructing Luminosity Functions and Schechter Fitting}
\label{sec:make_lfs}

Luminosity functions of our samples are measured via the commonly used $V_{max}$ estimator:
\begin{eqnarray}
	\Phi(L) = \frac{1}{\Delta \log_{10} L} \sum_{i} \frac{1}{C(L_i) V_{max,i}}
\end{eqnarray}
where $\Delta\log_{10} L$ is the bin width in log-space, $C(L_i)$ is the completeness correction factor for an individual galaxy with line luminosity $L$ from \S\ref{sec:completeness}, and $V_{max}$ is the survey comoving volume for the $i^\textrm{th}$ galaxy in the bin. Since our LAGER DECam image is a single pointing, our survey is homogeneous in line luminosity depth such that no image-to-image calculations for the comoving volume is required. We measure the comoving volume based on our $2.4$ deg$^2$ survey size\footnotemark and use the redshift range of our samples (based on the FWHM of the narrowband filter) as the limits of our measurement. Note that this approach assumes a top-hat filter and we take this effect into account in our filter profile corrections as described in \S \ref{sec:filter_volume_corr}. We assume Poisson errors for all our luminosity function bins and incorporate 20 percent error in quadrature from the completeness correction measurements, as described in \S\ref{sec:completeness}.

Since we are binning our measurements in log-space, we use the log-form of the Schechter function to fit our measurements of the luminosity function. This is described as:
\begin{eqnarray}
\centering
\Phi(L)\mathrm{d}L = \phi^\star \ln{10}\ \Bigg(\frac{L}{L^\star}\Bigg)^{1+\alpha} e^{-(L/L^\star)} \mathrm{d} \log_{10}L
\end{eqnarray}
where $\phi^\star$ and $L^\star$ are the characteristic number density and luminosities, respectively, $\alpha$ is the faint-end slope, and an added factor of $\ln{10}$ is included due to the log-form nature of the Schechter function. All our measurements are fitted using a MCMC approach along with a Metropolis-Hastings sampling with 50000 iterations. This allows us to investigate the probability distribution functions of our measured Schechter parameters and take into account the correlation between the three parameters.

\footnotetext{Although our survey is $\sim 3$ deg$^2$ in size, we are limited to the 2.4 deg$^2$ survey area covered by COSMOS2015 \citep{Laigle2016}, as shown in Figure \ref{fig:LAGER_footprint}. \citet{Laigle2016} cites the area as 2 deg$^2$ and is shown in {\it red} in Figure \ref{fig:LAGER_footprint} in their analyses, while the COSMOS2015 catalog includes sources in the outer 0.4 deg$^2$ region that is enclosed in the {\it blue} region in Figure \ref{fig:LAGER_footprint}. We include sources within this region in our study to maximize our survey area to 2.4 deg$^2$.}

\section{Luminosity Functions}
\label{sec:LFs}

\begin{figure*}
	\centering
	\includegraphics[width=\textwidth]{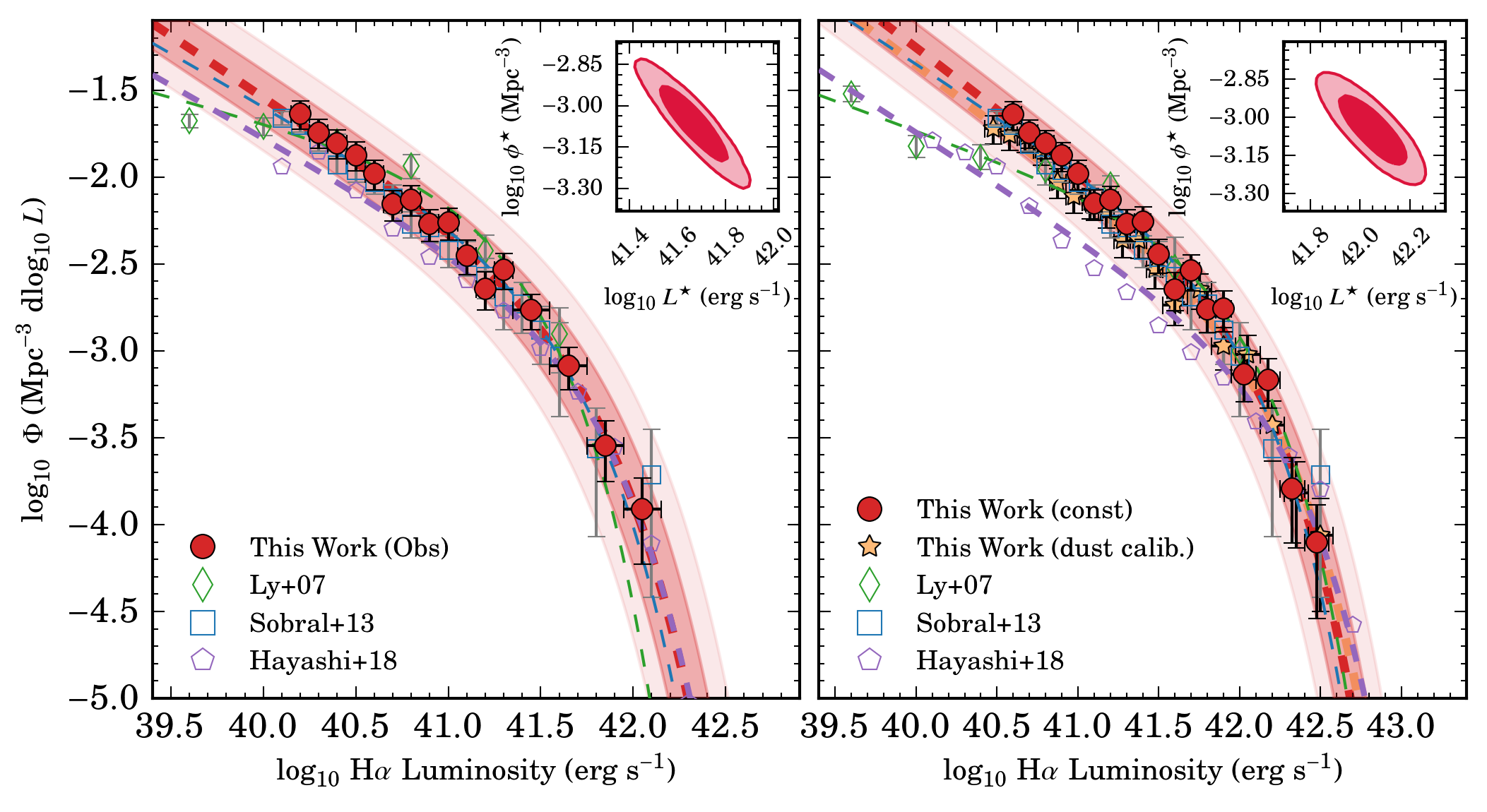}
	\caption{The $z = 0.47$ \ha~luminosity function with the best-fit based on $\alpha = -1.75$ fixed. The {\it left} panel shows our observed luminosity function and the {\it right} panel shows our luminosity functions when dust correcting using \aha~$=1$ mag ({\it red circles}) and our own empirical calibration ({\it orange stars}). We find excellent agreement between our observed luminosity functions and $z \sim 0.4$ literature measurements \citep{Ly2007,Sobral2013,Hayashi2018} for all luminosities probed. Errors for the \citet{Hayashi2018} LF measurements were not tabulated in their paper for use in this study. We find that the \citet{Ly2007} LF finds a shallower faint-end slope of $\alpha = -1.28$ as seen for luminosities fainter than $10^{40}$ erg s$^{-1}$, although we note their survey is 10 times smaller in areal coverage such that their results may be driven by cosmic variance. We find our LFs based on the two cases of dust corrections to be in agreement suggesting that \aha$= 1$ mag is a reasonable approximation to describe our \ha~samples. Our LFs are in strong agreement with \citet{Ly2007} and \citet{Sobral2013}, while the recent HSC survey of \citet{Hayashi2018} is finding slightly lower number densities.} 
	\label{fig:HA_LF}
\end{figure*}

\subsection{\ha~Luminosity Function}
\label{sec:HA_LF}

We present our $z = 0.47$ \ha~luminosity function in Figure \ref{fig:HA_LF} with the Schechter parameters shown in Tables \ref{table:LF_free} and \ref{table:LF_fixed}. To reduce the degeneracy effects of the three different Schechter parameters, we fix $\alpha$ to $-1.75$ which is consistent with $\alpha = -1.77^{+0.12}_{-0.11}$ when we treat the faint-end slope as a free parameter. We also include the LF measurements for which all 3 parameters are free in Table \ref{table:LF_free}. Figure \ref{fig:HA_LF} also includes as {\it dark} and {\it light} shaded regions the $1\sigma$ and $2\sigma$ confidence regions based on our MCMC fitting.

We find a best-fit observed luminosity function of \pstar$ = 10^{-3.16\pm0.09}$ Mpc$^{-3}$ and \lstar$ = 10^{41.72\pm0.09}$ erg s$^{-1}$ with $\alpha = -1.75$ fixed. The observed luminosity function is found to be in strong agreement with previous \ha~surveys of varying areal size and depth \citep{Ly2007,Sobral2013,Hayashi2018}. As our approach is very similar to the HiZELS survey \citep{Sobral2013}, it is not surprising that we strongly agree with their \ha~LF assessment. Although HiZELS covers 2 deg$^2$ with an \ha~volume of $8.8 \times 10^{4}$ Mpc$^{-3}$ roughly evenly split between the COSMOS and UDS fields, our measurements are based on a fixed $\sim 2.4$ deg$^2$ coverage in COSMOS with an \ha~volume of $11.2 \times 10^{4}$ Mpc$^{-3}$. The larger volume and similar depth allows us to have a better handle on both the bright- and faint-end.  For example, HiZELS observes 9 $L > L^\star$ \ha~emitters in comparison to our 18 $L > L^\star$ \ha~emitters in LAGER, such that we can better constrain the bright-end.

Our observed LF is also in agreement with the 0.24 deg$^2$ SDF measurement of \citet{Ly2007} within the luminosity range for which we observe. Below our 30\% completeness limit of $10^{40.14}$ erg s$^{-1}$, \citet{Ly2007} observes lower number densities with a faint-end slope of $\alpha = -1.28\pm0.07$, significantly shallower in comparison to ours as well as other narrowband surveys. We note that their samples have a rest-frame EW cut of $\sim 11$\AA, which is significantly lower than our $\sim 35$\AA~cut, such that they are sensitive to low luminosity systems, but also can have a higher risk in contamination due to stellar continuum mimicking an emission line in the narrowband filter. Furthermore, the shallower slope could also be a result of cosmic variance as their survey is 10 times smaller than LAGER. Future, deep narrowband data is needed to assess the validity of a shallower slope within this line luminosity range. In comparison to the recent 16 deg$^2$ Hyper-SuprimeCam survey of \citet{Hayashi2018}, we find their LF to be in agreement with our measurements, with the most significant deviation ($> 1\sigma$) found at their faintest line luminosities.

In the {\it right} panel of Figure \ref{fig:HA_LF}, we show the dust-corrected \ha~luminosity function for the case of a constant \aha$= 1$ mag ({\it red circles}) and when using our empirical dust calibration ({\it orange stars}) defined in \S\ref{sec:dust_calib}. Comparing both cases shows a strong agreement between the two luminosity functions. This suggests that assuming a constant dust extinction is representative of \ha~emitters as a whole population, as suggested by previous \ha~studies (e.g., \citealt{Ibar2013,Sobral2016b}). This is also shown by the agreement between our dust calibration-corrected luminosity and the \ha~LF of \citet{Sobral2013}, which also assumes a constant \aha$=1$ mag.

In regards to their Schechter fits, we find a best-fit luminosity function of \pstar$ = 10^{-3.14\pm0.09}$ Mpc$^{-3}$ and \lstar$ = 10^{42.10\pm0.09}$ erg s$^{-1}$ using our constant dust correction and \pstar$ = 10^{-3.32\pm0.10}$ Mpc$^{-3}$ and \lstar$ = 10^{42.19\pm0.11}$ erg s$^{-1}$ using our calibration with both measurements based on $\alpha = -1.75$ fixed. It is not surprising that the measured faint-end slopes between the observed, $\alpha = -1.77^{+0.12}_{-0.11}$, and \aha$=1$ mag corrected, $\alpha = -1.78^{+0.11}_{-0.10}$, luminosity functions are the same, since the dust-corrected LF is a rescaling in \lstar~of the observed LF. Interestingly, we measure a $\alpha = -1.77^{+0.10}_{-0.09}$ for the dust calibration-corrected LF, which is the same as the other two \ha~LF measurements. This is surprising as each \ha~emitter in the sample is dust-corrected differently based on their rest-frame $g - r$ colors such that the faint-end slope should not necessarily be the same. This would suggest that our \ha~luminosity functions are not sensitive to second-order dust corrections (e.g., non-constant dust extinction corrections), assuming that such a dust calibration (based on local measurements) is tested against observed \aha~measurements, as shown in Figure \ref{fig:dust_calib}.

\begin{figure*}
	\centering
	\includegraphics[width=\textwidth]{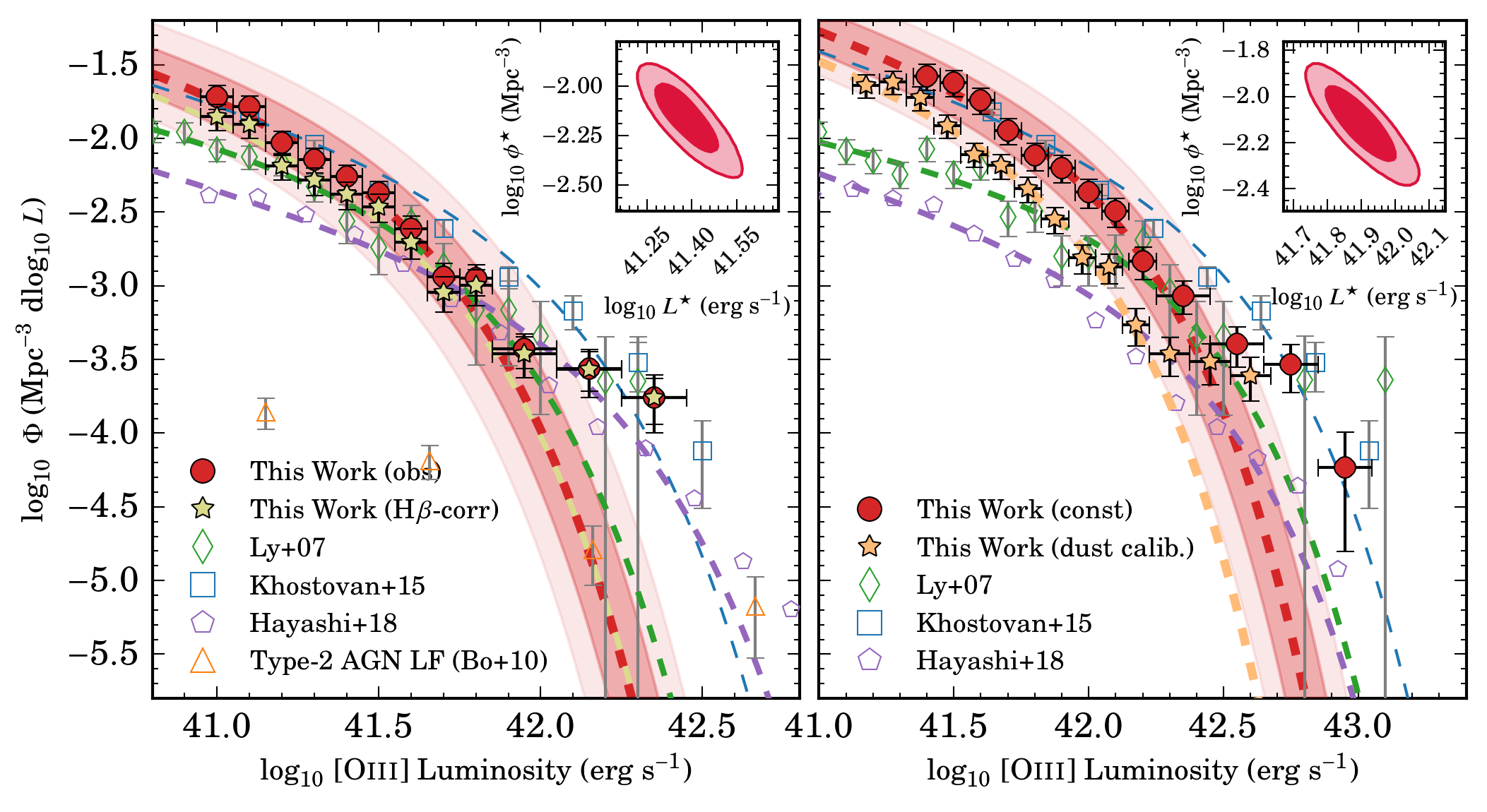}
	\caption{The $z = 0.93$ \oiii~luminosity function with the best-fit based on $\alpha = -1.6$ fixed. We show the case of our observed LF in the {\it left} panel as {\it red circles} with the $1\sigma$ and $2\sigma$ regions as {\it dark} and {\it light} shades of red, respectively. We also include the \oiii~LF corrected for \hb~contribution shown as {\it light green} stars, where we see no significant difference in comparison to our observed \oiii~LF. This suggests that the amount of \hb~emitters in our samples is negligible. Included are $z = 0.84$ measurements from the literature, where we find our LF is in agreement with \citet{Ly2007} and only in agreement with \citet{Hayashi2018} at bright line luminosities. We note that errors for the \citet{Hayashi2018} LF measurements are missing as they were not tabulated in their paper. A change in the LF is seen at luminosities $> 10^{41.9}$ erg s$^{-1}$, which may be due to AGN contamination. The {\it right panel} shows our dust-corrected LFs where we find that our two dust correction cases are not in agreement. Furthermore, we do not see any strong agreement with the literature, which highlights the varying dust corrections applied and also suggests further investigation of the dust properties of \oiii~emitters is needed.}
	\label{fig:O3_LF}
\end{figure*}

We find that our \ha~dust-corrected luminosity functions agree with \citet{Ly2007} down to $10^{40.5}$ erg s$^{-1}$ and \citet{Hayashi2018} down to $10^{41.5}$ erg s$^{-1}$. The discrepancy at fainter line luminosities shows the affects of different dust correction assumptions. In the case of \citet{Ly2007}, they use the luminosity-dependent dust correction relation of \citet{Hopkins2001}, which assumes a \citet{Cardelli1989} dust attenuation curve. This would suggest that their dust correction is still representative of a constant \aha$=1$ mag. The discrepancy below $10^{40.5}$ erg s$^{-1}$ is most likely due to the same argument made for the case of the observed LFs, where the survey area is small and the low EW cut could be contaminating their samples.

The \citet{Hayashi2018} LF is found to be systematically lower for luminosities $< 10^{41.5}$ erg s$^{-1}$ and uses SDSS DR7-selected \ha~emitters along with their Balmer decrements as dependent on observed line luminosity and stellar mass to dust correct their samples. This is a somewhat similar approach to our empirical dust calibration such that the disagreement highlights systematically different dust corrections applied per line luminosity.

\subsection{\oiii~Luminosity Function}
\label{sec:O3_LF}

Figure \ref{fig:O3_LF} shows our $z = 0.93$ \oiii~luminosity function with the best-fit Schechter parameters presented in Tables \ref{table:LF_free} and \ref{table:LF_fixed}. In comparison to our \ha~luminosity functions, we can not strongly constrain the faint-end slope where we find $\alpha = -1.57^{+0.35}_{-0.30}$, $-1.73^{+0.29}_{-0.19}$, and $-1.79^{+0.29}_{-0.15}$ for our observed, constant \aha~, and our dust calibration correction, respectively. Given this issue, we fit the Schechter model assuming a constant $\alpha = -1.60$, which is consistent when setting $\alpha$ as a free parameter. These fits are shown as dashed lines in Figure \ref{fig:O3_LF} with the best-fit \pstar~and \lstar~ shown in Table \ref{table:LF_fixed}. 

We find our observed luminosity function is best represented by a Schechter function up to $10^{41.90}$ erg s$^{-1}$ with $\phi^\star = 10^{-2.16^{+0.10}_{-0.12}}$ Mpc$^{-3}$ and $L^\star = 10^{41.38^{+0.07}_{-0.06}}$ erg s$^{-1}$ with $\alpha = -1.60$ (fixed). At brighter luminosities, we notice a shallower decrease in the number densities with increasing \oiii~luminosity. This may be due to the changing nature of the \oiii~emission line where it is mostly tracing AGN rather than star formation activity (e.g., \citealt{Kauffmann2003,Heckman2005,Kauffmann2009,Wild2010,Azadi2017}). Using the full range of luminosity bins in Figure \ref{fig:O3_LF}, we find a best-fit Schechter function of $\phi^\star = 10^{-2.30\pm0.11}$ Mpc$^{-3}$ and $L^\star = 10^{41.48\pm0.06}$ erg s$^{-1}$ with $\alpha = -1.60$ (fixed). We find a reduced $\chi^2 = 2.30$, while constraining the fit to luminosities fainter than $10^{41.90}$ erg s$^{-1}$ results in a reduced $\chi^2 = 0.55$ providing a better fit. Offsets from a Schechter function due to AGN contribution in the bright-end has also been observed in other \ha, \oiii, \oii, and \lya~studies (e.g., \citealt{Matthee2017,Wold2017,Sobral2018b}). If the \oiii~emission is driven by AGN activity in the bright-end, then comparing the observed Schechter fit to the bright-end suggests 95 - 100 percent of \oiii~emitters at $L_\textrm{\oiii} > 10^{41.90}$ erg s$^{-1}$ ($L_\textrm{\oiii} > 3L^\star$) are AGN. This is similar to recent results of other ELG studies that find $> L^\star$ populations dominated by AGN \citep{Sobral2016,Matthee2017,Sobral2018}.

We test the possibility of AGN contamination being the cause of the excess in the bright-end by comparing the total luminosity densities of $L > 10^{41.90}$ erg s$^{-1}$ emitters to the total sample. This is done by integrating the observed $\Phi(L)$ bins ({\it red circles} in Figure \ref{fig:O3_LF}) weighted by line luminosity (e.g., $\int L \Phi(L) \mathrm{d}L$). We find that for $L > 10^{41.90}$ erg s$^{-1}$, the luminosity density is $10^{38.14}$ erg s$^{-1}$ Mpc$^{-3}$ in comparison to $10^{39.07}$ erg s$^{-1}$ Mpc$^{-3}$ for the luminosity range between $10^{41}$ erg s$^{-1}$ to $10^{42.35}$ erg s$^{-1}$ (covering the full range of our sample). This corresponds to a bright-end contribution of $\sim 12\%$, which is similar to the $13\%$ AGN contamination measured in \S\ref{sec:AGNs}. This suggests that if the bright-end is dominated by AGN at $L > 3 L^\star$, then the fraction of light that the bright-end contributes is consistent with the expected AGN contamination of the sample. This does not mean that every individual bright-end \oiii~emitters are AGN. We only point out the consistency between expected AGN contamination and the bright-end contribution.
	
Included in Figure \ref{fig:O3_LF} is the $0.5 < z < 0.92$ type-2 AGN $z$COSMOS luminosity function of \citet{Bongiorno2010}. We find that for our faintest \oiii~emitters, a contribution of $\sim 1\%$ comes from type-2 AGN and about $10\%$ by $10^{41.90}$ erg s$^{-1}$. \citet{Matthee2017} finds a $20\%$ X-ray AGN fraction for their brightest $z = 0.8$ \oiii~emitters. We note that the AGN fraction cited is for a specific AGN selection and omits contribution from optical-, infrared-, and radio-selected AGN, which could contribute more in the bright-end. Spectroscopic follow-up is required to properly assess the level of AGN contamination in the bright-end of our luminosity function. That being given, further \oiii~measurements mentioned in this paper are based on the luminosity functions fitted with the Schechter form up to $10^{41.9}$, $10^{42.2}$, and $10^{42.3}$ erg s$^{-1}$ for our observed, dust-calibration corrected, and constant \aha~corrected cases, respectively.

As noted in \S\ref{sec:contrib_hbeta}, our narrowband filter can properly separate between \hb~and \oiii~emitters, but our photo-$z$ and color-color selections lack the resolution to decouple the two populations. Our spectroscopic matches show that \hb~emitters contribute $< 5\%$ to our total \oiii-selected sample, although we note there is a selection bias to brighter line emitters such that our measured \hb~contribution may not be representative of the whole sample. Previous studies at a similar redshift find a $10 - 20\%$ \hb~contribution at fainter observed line luminosities \citep{Sobral2015,Khostovan2015,Khostovan2016}.

We correct our observed \oiii~luminosity function for \hb~contribution by using the \ha~luminosity function as a proxy. We start by redshift projecting the $z = 0.84$ \ha~luminosity function of \citet{Sobral2013} given their measured redshift evolution of \pstar~and \lstar~with a faint-end slope that is fixed to $\alpha = -1.6$. Using the intrinsic \ha/\hb~line ratio of $2.86$ (assuming case B recombination) and removing the \aha$=1$ mag dust correction assumed in \citet{Sobral2013}, we predict an \hb~luminosity function of \pstar$=10^{-2.60}$ Mpc$^{-3}$ and \lstar$=10^{41.29}$ erg s$^{-1}$. We then subtract the \hb~number densities from our observed \oiii~luminosity function for each given observed line luminosity.

The \hb-corrected \oiii~luminosity function is shown as {\it green stars} in Figure \ref{fig:O3_LF} with the best-fit Schechter parameters as $\phi^\star = 10^{-2.33\pm0.11}$ Mpc$^{-3}$ and $L^\star = 10^{41.41\pm0.07}$ erg s$^{-1}$ for a fixed $\alpha = -1.6$. We find that there is no statistically significant effect such that our samples are primarily tracing \oiii~emitters. The number densities are systematically lower at fainter observed luminosities, but they are still within $1\sigma$ agreement to our observed \oiii~LF. This suggests that the contribution of \hb~emitters is not statistically significant in our samples.

We find our observed luminosity function is in reasonable agreement with \citet{Ly2007} for line luminosities $> 10^{41.5}$ erg s$^{-1}$, while they report lower number densities towards fainter line luminosities. \citet{Hayashi2018} report even lower number densities in the faint-end. The discrepancy may be due to the effects of two possible overdense regions at $z = 0.93$ identified by \citet{Scoville2007}. Since \citet{Hayashi2018} covers 16 deg$^2$, it is expected that overdensity effects would be washed out by field emitters in the larger area. Matching with the publicly available, mass-complete COSMOS environment catalog \citep{Darvish2017}, we observe filamentary structure and overdense regions within our field and find faint \oiii~emitters being the primary members. This would suggest that part of the significant number density differences is attributed to large-scale structure effects, although second-order effects (e.g., source selection, volume assessment) may also contribute to the discrepancies.
	
In comparison to \citet{Khostovan2015}, we find agreement only in their faintest line luminosities. We report lower number densities at brighter line luminosities, which could suggest the \citet{Khostovan2015} sample may be more AGN-dominated for \oiii~luminosities $>10^{41.9}$ erg s$^{-1}$. It becomes even more evident that AGN play a more important role at bright line luminosities, such that the \citet{Hayashi2018} should be 100\% dominated by type-2 AGN by 10$^{42.5}$ erg s$^{-1}$ when compared to \citet{Bongiorno2010}. 

\begin{figure*}
	\centering
	\includegraphics[width=\textwidth]{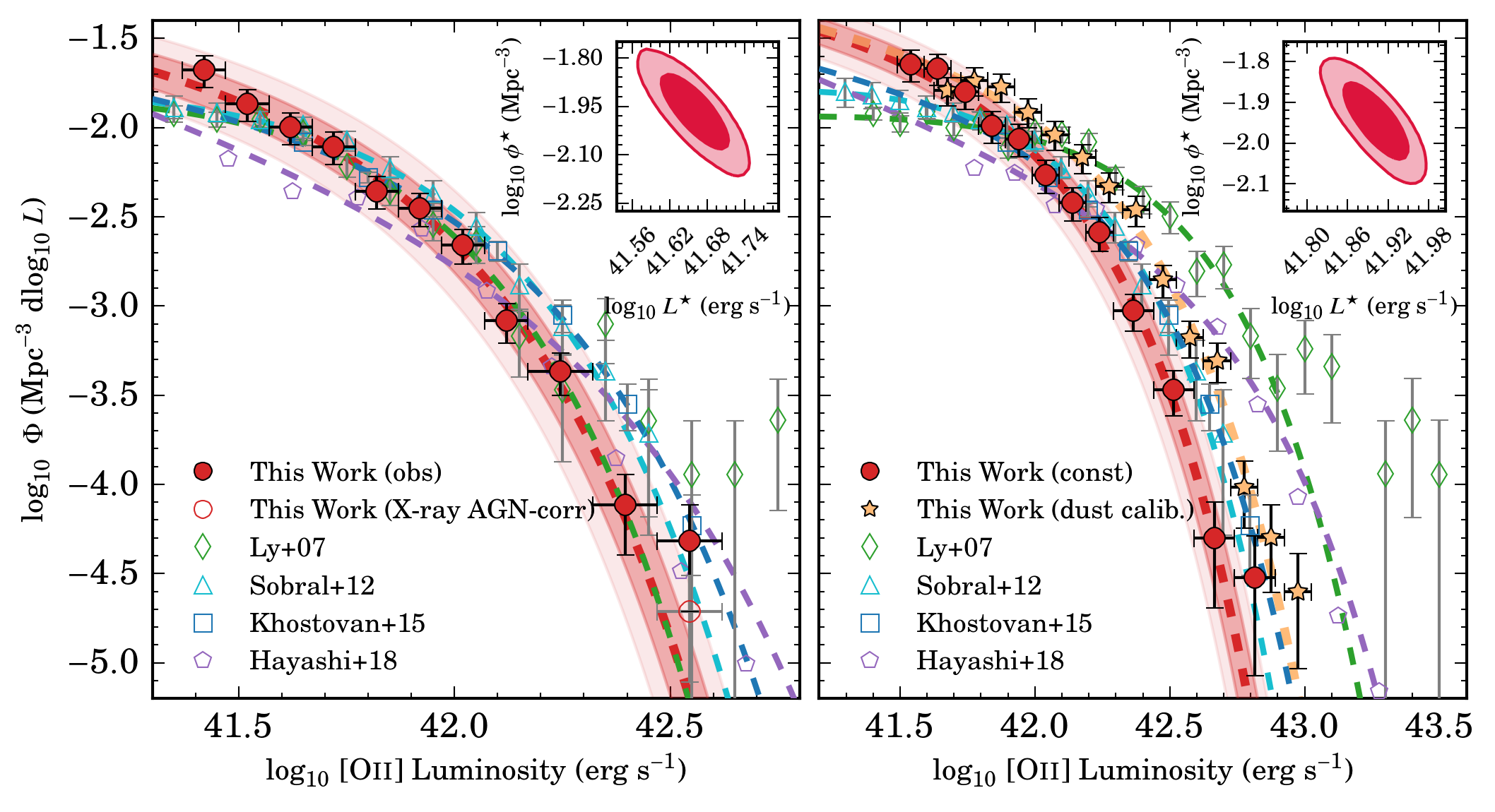}
	\caption{The $z = 1.59$ \oii~luminosity function with the best-fit based on $\alpha = -1.3$ fixed. The {\it left panel} shows our observed luminosity function along with $z \sim 1.5$ measurements from the literature. We find our LF to be in agreement with past assessments of the \oii~LF, with the most significant disagreement arising in the faint-end where \citet{Hayashi2018} measures lower number densities. The missing errors for \citet{Hayashi2018} measurements are due to the lack of tabulated LF measurements in their paper. We also find 60\% (3/5) of \oii~emitters in the brightest luminosity bin to be X-ray detected AGNs. Correcting for this contribution reduces the number density to $10^{-4.7}$ Mpc$^{-3}$, better matching our Schechter fit. This highlights the importance of understanding the AGN contribution in the bright-end for ELG samples. The {\it right panel} shows our dust corrected luminosity functions. For the case of the constant dust correction, we assume \aha$=0.35$ mag as suggested by \citet{Hayashi2013}. We find that our LFs are shifted towards fainter line luminosities in comparison to the literature, which arises from different dust prescriptions.}
	\label{fig:O2_LF}
\end{figure*}

The {\it right panel} of Figure \ref{fig:O3_LF} shows our luminosity functions with a constant \aha$=1$ mag applied (assuming \citet{Calzetti2000} dust attenuation curve corresponds to $A_\textrm{\oiii} = 1.35$ mag; {\it red circles}) and our empirical dust calibration ({\it orange stars}). A best-fit of \pstar$ = 10^{-2.12\pm0.10}$ Mpc$^{-3}$ and \lstar$ = 10^{41.90^{+0.07}_{-0.06}}$ erg s$^{-1}$ is measured for our constant \aha~dust corrected LF and \pstar$ = 10^{-2.20^{+0.09}_{-0.08}}$ Mpc$^{-3}$ and \lstar$ = 10^{41.73\pm0.05}$ erg s$^{-1}$ is measured for our dust calibration-corrected LF. We find that our dust calibration-corrected LF is systematically fainter than our constant-corrected LF. We caution that our calibration was tested against our $z = 0.47$ \ha~samples and may not be representative of our $z = 0.93$ \oiii~samples. 

We find an agreement only with the bright-end of \citet{Ly2007} LF and the \citet{Hayashi2018} LF being systematically lower in number densities, although this is partly attributed to the wider area as described above. The main cause of the discrepancy can be attributed to the different dust correction assumptions. \citet{Ly2007} uses the \citet{Hopkins2001} \ha~luminosity-dependent calibration with \oiii/\ha~line ratios measured using an \oiii/\ha--$M_B$ relation (calibrated using their NB704 and NB921 samples; observes \oiii~and \ha~at $z \sim 0.4$, respectively). \citet{Hayashi2018} applies dust corrections by assuming the line ratios as dependent on stellar mass and observed line luminosity of SDSS DR7-selected emitters. Overall, the role of dust is found to be more complex than what was seen for our \ha~LFs and requires a thorough and detailed investigation.

\subsection{\oii~Luminosity Function}
\label{sec:O2_LF}

We present our $z = 1.59$ \oii~luminosity functions in Figure \ref{fig:O2_LF} with the best-fit Schechter parameters shown in Tables \ref{table:LF_free} and \ref{table:LF_fixed}. Although we measure the faint-end slopes to be $\alpha = -1.58^{+0.30}_{-0.27}$, $-1.47^{+0.23}_{-0.23}$, and $-1.21^{+0.15}_{-0.20}$ for our observed, constant \aha~corrected, and dust calibration corrected LFs, we note that we do not provide strong constraints and therefore set $\alpha = -1.3$ and show the corresponding LFs in Figure \ref{fig:O2_LF}.

The observed \oii~luminosity function is measured to have a best-fit of \pstar$ = 10^{-1.97\pm0.07}$ Mpc$^{-3}$ and \lstar$ = 10^{41.66\pm0.03}$ erg s$^{-1}$ with $\alpha = -1.3$ fixed. We observe a slight excess in bright sources above $>10^{42.55}$ erg s$^{-1}$ which we attribute to a higher AGN fraction. For the \oii~luminosities between $10^{42.4 - 42.6}$ erg s$^{-1}$, we found 3 X-ray detections (see \S\ref{sec:AGNs}) out of a total of 5 \oii~emitters observed within this luminosity range, suggesting a 60\% X-ray AGN fraction. Correcting for this by reducing the number density by 60\% results in a number density of $10^{-4.7}$ Mpc$^{-3}$ dlog$_{10}L^{-1}$, which brings it in better agreement with our best-fit Schechter model.

Our \oii~luminosity function does not probe as deep as \citet{Ly2007} and \citet{Sobral2012}, but we note the two respective studies are based on small areas (0.24 deg$^2$ and 0.67 deg$^2$, respectively). Our measurement of the \oii~LF is about 0.1 dex deeper than the 16 deg$^2$ HSC survey \citep{Hayashi2018}.

We find our observed \oii~luminosity function to be mostly consistent with measurements from the literature. There is a strong agreement in the bright-end between our \oii~number densities and those of \citet{Hayashi2018}, although we find their number densities to be lower for luminosities $< 10^{41.8}$ erg s$^{-1}$. At the faint-end, \citet{Ly2007} and \citet{Sobral2012} report number densities $\sim 10^{-1.95}$ Mpc$^{-3}$ d$\log_{10}L^{-1}$, which is $\sim 0.25$ dex lower than our measurement. At the bright-end, we find the \citet{Ly2007} number densities to be systematically higher than our measurements, although their number statistics are poor in this regime resulting in large error bars.

In comparison to the 2 deg$^2$ HiZELS measurement of \citet{Khostovan2015}, we probe 0.2 dex deeper in line luminosity. We are in agreement for line luminosities $< 10^{42.1}$ erg s$^{-1}$ as well as for our brightest luminosity bin. For the rest of the bright-end, we find the \citet{Khostovan2015} number densities to be systematically higher, although within 2$\sigma$ agreement based on our Schechter fit.

The {\it right panel} of Figure \ref{fig:O2_LF} shows our dust-corrected luminosity functions. For our \ha~and \oiii~samples, we assumed a constant \aha$=1$ mag. \citet{Hayashi2013} investigated the dust properties and \oii/\ha~ratios of $z = 1.47$ \oii~emitters using a double blind \ha~and \oii~narrowband survey. They found that for the luminosity range between $10^{41.2 - 42.8}$ erg s$^{-1}$, \oii~emitters selected via narrowband surveys are dust-poor and are better represented by \aha~$=0.35$ mag. They also find evidence from comparison to local SDSS measurements that \oii~emitters with faint observed luminosities tend to be dustier than bright emitters. This would suggest an anti-correlation between \aha~and observed line luminosities. Similar studies have found an anti-correlation between dust attenuation and \oii/\ha~line ratios (e.g., \citealt{Moustakas2006}). Given this result, we show our dust-corrected \oii~luminosity function for the case that \aha$=0.35$ mag (corresponding to $A_\textrm{\oii} = 0.62$ mag assuming \citet{Calzetti2000} dust attenuation curve) and also for the case of our empirical dust calibration. We note that the median dust correction applied using our calibration is \aha$=0.55\pm0.12$ mag, which is higher than our constant dust correction. 

We find our constant \aha~corrected LF is best fit by \pstar$ = 10^{-1.95\pm0.06}$ Mpc$^{-3}$ and \lstar$ = 10^{41.90\pm0.03}$ erg s$^{-1}$ and our dust calibration corrected LF by \pstar$ = 10^{-2.02\pm0.05}$ Mpc$^{-3}$ and \lstar$ = 10^{42.10\pm0.03}$ erg s$^{-1}$ with $\alpha = -1.3$ fixed in both measurements. We observe a systematic shift between the \aha~corrected and calibration corrected LFs due to the median dust correction using our calibration being \aha$=0.55\pm0.12$ mag, which is higher than our constant \aha$=0.35$ mag correction. In comparison to measurements from the literature, we find that previous results also are in disagreement with one another, which highlights the effects of varying dust correction prescriptions. Spectroscopic follow-up of our sample is needed so that we can measure Balmer decrements and thus robustly investigate the dust properties of \oii~emitters.

\begin{figure*}
	\centering
	\includegraphics[width=\textwidth]{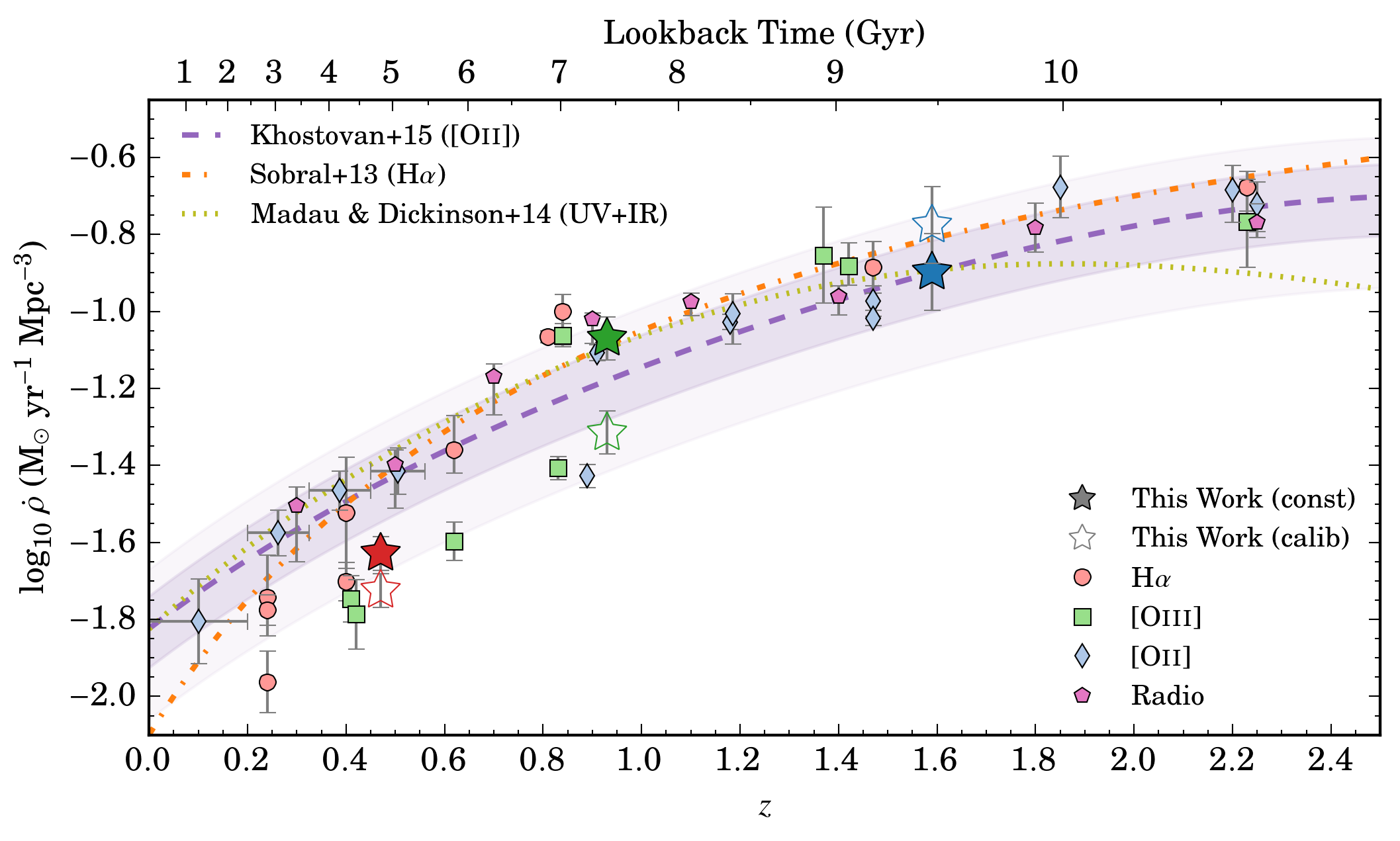}
	\caption{The cosmic star-formation rate density (SFRD) evolution up to $z \sim 2.5$. Our SFRDs with the constant dust correction are shown as {\it filled stars} and those corrected using our empirical calibration are shown as {\it empty stars}. All our SFRDs are corrected for AGN contamination with the amount of correction added to the errors in quadrature. We include previous \ha~(\citealt{Ly2007,Shioya2008,Sobral2013,Sobral2015,Stroe2015}, Harish et al., in submitted), \oiii~\citep{Ly2007,Khostovan2015,Sobral2015}, \oii~\citep{Ly2007,Takahashi2007,Bayliss2011,Ciardullo2013,Khostovan2015,Sobral2015}, and the 1.4 Hz radio-stacked measurements of \citet{Karim2011}. We find that, regardless of emission line selection, our SFRDs are in agreement with previous measurements. Only the \oiii~and \oii~SFRDs (constant \aha) are found to be in agreement with \citet{Karim2011}, suggesting that the dust prescription is representative of the sample, while the \ha~SFRD is lower, similar to previous studies. We also include the \oii~\citep{Khostovan2015}, \ha~\citep{Sobral2013}, and UV+IR \citep{Madau2014} parameterizations of the cosmic SFRD evolution and find our samples are, generally, in agreement.}
	\label{fig:SFRD}
\end{figure*}

\section{Cosmic Star Formation History}
\label{sec:SFRDs}

One of the fundamental properties in galaxy evolution physics is the evolution of the cosmic star formation rate density, which quantifies the amount of star formation activity within a given comoving volume at a specific epoch in cosmic time. Here exists the effects of all physical properties that govern/drive star formation activity throughout the Universe. Careful, consistent, and robust measurements on the cosmic star formation history are then needed to properly constrain this observable for future analysis of the underlying physics.

Previous compilations show large scatter arising from the different indicators tracing varying timescales of star formation activity, as well as different dust correction prescriptions applied \citep{Hopkins2004,Hopkins2006,Madau2014}. A major advantage of using emission line samples to investigate the cosmic star-formation history is the consistency of tracing similar star-formation timescales ($\sim 10$ Myr; observing massive, short-lived, $O$ and $B$ stars).

We use our luminosity functions as shown and discussed in \S\ref{sec:HA_LF} -- \ref{sec:O2_LF} to measure the star formation rate densities. This is done by integrating the luminosity functions for all luminosities, which is defined as:
\begin{eqnarray}
	\centering
	\rho_L = \int\displaylimits_{0}^{\infty} L\ \Phi(L) dL = \phi^\star L^\star \Gamma(2+\alpha)
\end{eqnarray}
where $\rho_L$ is the luminosity density in erg s$^{-1}$ Mpc$^{-3}$. Although there is an expected threshold in luminosity for which one could have a galaxy, such a `turnover' in the luminosity functions have yet to be observed. For example, the $z \sim 2$ ultra-faint, UV lensing studying of \citet{Alavi2014} observed sources to M$_\textrm{UV} \sim -12$ mag and reported no turnover in the luminosity function. Therefore, we measure our luminosity and star formation rate densities for all line luminosities.

\begin{table*}
	\centering
	\caption{Star Formation Rate Densities (SFRDs) for each emission line sample with our two dust correction applications. Note that we apply a constant \aha$=0.35$ mag to our \oii~measurements as suggested by \citet{Hayashi2013}. For each dust correction case, we show the luminosity density, $\rho_L$, the SFRD integrated for all luminosities, $\rho_\textrm{SFR}$, and the SFRD integrated for all luminosities with the addition of an AGN correction as described in \S\ref{sec:AGNs}. We also apply the AGN correction in quadrature to our errors. Comparing the two cases, we only see an agreement for our \ha~and \oii~samples. The \oiii~sample shows a 2$\sigma$ difference between the two dust cases, suggesting that our measurement is dominated by systematics arising from the dust correction.}
	\begin{tabular*}{\textwidth}{@{\extracolsep{\fill}} c c c c c c c}
		
		\hline
		& \multicolumn{3}{c}{Constant $A_\textrm{\ha}$} & \multicolumn{3}{c}{Dust Calibration} \\
		\cline{2-4} \cline{5-7}
		Line & $\log_{10} \rho_L$ & $\log_{10} \rho_\textrm{SFR}$ & $\log_{10} \rho_\textrm{SFR,AGN-corr}$ & $\log_{10} \rho_L$ & $\log_{10} \rho_\textrm{SFR}$ & $\log_{10} \rho_\textrm{SFR,AGN-corr}$\\
		& (erg s$^{-1}$ Mpc$^{-3}$) & (M$_\odot$ yr$^{-1}$ Mpc$^{-3}$) & (M$_\odot$ yr$^{-1}$ Mpc$^{-3}$) & (erg s$^{-1}$ Mpc$^{-3}$) & (M$_\odot$ yr$^{-1}$ Mpc$^{-3}$) & (M$_\odot$ yr$^{-1}$ Mpc$^{-3}$)   \\
		\hline
		\ha & $39.52^{+0.03}_{-0.03}$ & $-1.58^{+0.03}_{-0.03}$ & $-1.63^{+0.04}_{-0.04}$ & $39.42^{+0.03}_{-0.03}$ & $-1.68^{+0.03}_{-0.03}$ & $-1.72^{+0.04}_{-0.04}$\\
		\oiii & $40.12^{+0.05}_{-0.04}$ & $-1.01^{+0.05}_{-0.04}$ & $-1.07^{+0.06}_{-0.06}$ & $39.88^{+0.04}_{-0.04}$ & $-1.25^{+0.04}_{-0.04}$ & $-1.31^{+0.06}_{-0.06}$\\
		\oii & $40.07^{+0.04}_{-0.04}$ & $-0.78^{+0.04}_{-0.04}$ & $-0.90^{+0.10}_{-0.10}$ & $40.19^{+0.03}_{-0.04}$ & $-0.66^{+0.03}_{-0.04}$ & $-0.78^{+0.10}_{-0.10}$\\
		\hline
	\end{tabular*}
	\label{table:SFRD}
\end{table*}

The star-formation rate densities are determined by using $\rho_L$ and convolving it with the star formation calibrations:
\begin{eqnarray}
	\rho_\textrm{SFR}(\textrm{\ha}) = 7.9 \times 10^{-42} \rho_L\\
	\rho_\textrm{SFR}(\textrm{\oiii}) = 7.35 \times 10^{-42} \rho_L\\
	\rho_\textrm{SFR}(\textrm{\oii}) = 1.4 \times 10^{-41} \rho_L
\end{eqnarray}
where $\rho_\textrm{SFR}$ is the star-formation rate density (SFRD) in M$_\odot$ yr$^{-1}$ Mpc$^{-3}$. We use the \ha~and \oii~calibrations of \citet{Kennicutt1998} and the relation derived from \citet{Osterbrock2006} for the \oiii~calibration with the assumption of a Salpeter IMF.

Our measured luminosity densities and star-formation rated densities for the case of a constant dust correction and also when applying our empirical dust calibration are shown in Table \ref{table:SFRD}. We also include the AGN-corrected SFRDs, which take into account the 10\%, 13\%, and 23\% AGN contamination found in \S\ref{sec:AGNs} for our \ha, \oiii, and \oii~samples, respectively. This was done by reducing the SFRDs by the amount of AGN contamination and also adding the correction to the errors in quadrature.

Figure \ref{fig:SFRD} shows the cosmic star formation rate density evolution with our AGN-corrected SFRD measurements shown as {\it stars}. For the two dust cases, we show {\it filled stars} as our SFRDs based on a constant \aha~correction and {\it empty stars} based on our empirical dust calibration. We find the SFRDs of the two dust correction methods to be in agreement with one another for our \ha~and \oii~samples, although the latter has large error bars due to the 23\% AGN contamination correction. Our \oiii~SFRDs are in strong disagreement which arises from the different dust calibration applied. Overall, we see that the SFRD based on our samples alone show about a dex decrease in star formation activity over 4 Gyr~of cosmic time.

Included in Figure \ref{fig:SFRD} are literature measurements for samples selected based on \ha~(\citealt{Ly2007,Shioya2008,Sobral2013,Sobral2015,Stroe2015}, Harish et al., in prep), \oiii~\citep{Ly2007,Khostovan2015,Sobral2015}, \oii~\citep{Ly2007,Takahashi2007,Bayliss2011,Ciardullo2013,Khostovan2015,Sobral2015}, and the 1.4 Hz radio-stacked measurements of \citet{Karim2011}. We include the radio stack measurements as these are not susceptible to dust attenuation. In the case of different SFR calibrations, dust corrections, and IMF assumptions, we recompute the literature SFRDs using their measured luminosity functions and set the SFRDs to the same underlying assumptions in order to ensure a compatible comparison with our measurements.

Figure \ref{fig:SFRD} also shows the \ha~\citep{Sobral2013}, \oii~\citep{Khostovan2015}, and UV+FIR~\citep{Madau2014} model of the cosmic SFRD evolution. We find our samples are in general agreement with the \oii~determination of \citet{Khostovan2015}, while our \ha~SFRD is not in agreement with the \citet{Sobral2013} and \citet{Madau2014} model.

We find our constant \aha-corrected \oiii~and \oii~SFRDs to be consistent with radio observations, suggesting that the dust correction applied may be representative of the sample. We also find strong agreement between our \oiii~and \oii~SFRDs and the \ha, \oiii, and \oii~literature measurements. Specifically the agreement with \ha, which is a well-calibrated tracer of star formation activity, suggests that \oiii~and \oii~emitters can still be used as `good' tracers of star formation despite the caveats, such as metallicity effects. The strong agreement also shows that our \oiii~and \oii~samples fully trace the star formation activity at $z = 0.93$ and $1.59$, respectively.

Around $z \sim 0.2 - 0.4$, we note a scatter in the \ha~literature measurements. Our \ha~SFRD is within the scatter and we find that it traces $\sim 70$ percent of the total $z = 0.47$ cosmic star formation activity. Our \ha~SFRDs are also below the radio-stack measurements of \citet{Karim2011}, which could signify that our \ha~sample has a population of dust-obscured star-forming galaxies that are under-corrected for dust. A dust correction of \aha$\sim 1.5$ mag would be needed to bring our \ha~SFRD in agreement with radio measurements. This corresponds to stellar masses above $>10^{10.5}$ \msol~assuming the \citet{Garn2010} \aha-stellar mass relation. Such systems could be potential luminous/ultra-luminous infrared galaxies (LIRGs/ULIRGs) with high levels of dust obscuration (e.g., \citealt{Casey2014} and references therein), such that our \aha $\sim 1$ mag dust correction is an underestimation. We note caution as this interpretation assumes that the \ha~and radio trace the same population of star-forming galaxies, which may not be entirely accurate. Further investigation of the low$-z$ \ha~dust properties, as well as selection biases, is required to understand the origin of the scatter within this epoch of cosmic time.

\section{Conclusions}
\label{sec:conclusion}

We present our results from the 3 deg$^2$ CTIO/Blanco DECam imaging of the COSMOS field as part of the larger 24 deg$^2$ LAGER survey. We highlight the main results below:
\begin{enumerate}[leftmargin=*]
	\item We select a large sample of 1577 $z = 0.47$ \ha, 3933 $z = 0.93$ \oiii, and 5367 $z = 1.59$ \oii~emission line galaxies using a combination of spectroscopic confirmations, photometric redshifts, and color-colors. Our samples are one of the largest at their respective redshifts and cover comoving volumes of $(1 - 7) \times 10^5$ Mpc$^3$, which greatly reduces the effects of cosmic variance in our measurements.
	\item Contamination of our samples are measured to be $\sim 6\%$, $\sim 10\%$, and $19\%$ for our \ha, \oiii, and \oii~samples, respectively. The higher contamination in the \oii~samples is attributed to sources with strong 4000\AA~breaks. Our color-color selections marginally increase the total contamination, but significantly increase the total sample size.
	\item A total of 45 X-ray detections are found for all our emission line samples, which would suggest $< 1$\% contamination. Using an infrared-selection based on the 1.6$\micron$ bump and AGN power-law SEDs, we find 10\%, 13\%, and 23\% AGN contamination for our \ha, \oiii, and \oii~samples, respectively. We also use archival spectra to measure AGN fractions for our \ha~sample using the MEx diagnostic and find $\sim 5\%$ contamination.
	\item We measure our observed luminosity functions, correcting for completeness and the NB filter profile. Additionally, our \ha~LFs are also corrected for \nii~contamination. We initially fit our observed LFs with $\alpha$ as a free parameter and find $\alpha = -1.77^{+0.12}_{-0.11}$, $\alpha = -1.57^{+0.35}_{-0.30}$, and $\alpha = -1.58^{+0.30}_{-0.27}$ for \ha, \oiii, and \oii, respectively. We subsequently fit the LFs with $\alpha$ fixed to $-1.75$, $-1.60$, and $-1.30$ for \ha, \oiii, and \oii~respectively. \item We find our \ha~and \oii~LFs to be in agreement with those from the literature, while the \oiii~LFs are found to have higher number densities. Correcting for \hb~contribution still shows higher number densities in comparison to the literature. This excess may be due to possible overdense regions that are inflating the number of sources in the survey.
	\item An excess of $>10^{42}$ erg s$^{-1}$ \oiii~emitters are found and is suggested to be caused by AGN contamination above $L^\star$. Further spectroscopic follow-up is needed to quantify the AGN fractions of bright emission line galaxies.
	\item We apply two forms of dust-corrections to our LFs: a constant \aha$ = 1$ mag (\aha$ = 0.35$ mag for \oii) and our own rest-frame $(g - r)$ dust calibration based on SDSS DR12 spectra and tested against archival $z$COSMOS spectra of our $z = 0.47$ \ha~emitters. We find no significant difference between the two \ha~dust-corrected LFs while the \oiii~and \oii~LFs show systematic differences. In comparison to the literature, we find no strong agreements between the \oiii~dust-corrected LFs. This suggests a detailed investigation of the nature of dust in \oiii~emitters is needed.
	\item Star formation rate densities are measured for both dust-corrected cases and AGN corrections are also applied. We find our SFRDs are in agreement with other emission line SFRDs in the literature, although our samples are larger in size and cover a wider area.
	\item We find our \oiii~and \oii~samples fully trace the cosmic SFRD at their respective redshifts, while our $z = 0.47$ \ha~sample is found to trace $\sim 70$ percent of the cosmic SFRD.
	\item Comparison to radio-stacked SFRDs shows that our \oiii~and \oii~constant \aha~dust corrections are representative of the samples. We find $z < 0.5$ \ha~emitters (from LAGER and the literature) are typically 0.2 dex below the radio-stacked SFRDs. This suggests the possibility of a sub-population of dust-obscured, star-forming galaxies with \aha$ > 1.5$ mag within our samples for which we are underestimating their dust corrections. Such dust properties are consistent with LIRGs and ULIRGs.
\end{enumerate}

Our emission line samples presented in this paper are the first from the LAGER survey. In total, LAGER will encompass 8 fields with a 3 deg$^2$ coverage per each field resulting in a combined survey area of 24 deg$^2$ corresponding to comoving volumes of 1.1, 3.4, and $6.5 \times 10^{6}$ Mpc$^{3}$ for our \ha, \oiii, and \oii~emitters, respectively. Using our observed luminosity functions and assuming similar 30\% completeness limits, we expect the full LAGER survey to have $\sim 13000$, $29000$, and $53000$ \ha, \oiii, and \oii~emitters, respectively. Upon completion, this would be the largest and deepest narrowband survey in the field and would present robust constraints on our understanding of star-forming galaxies.

\section*{Acknowledgments}
We thank the anonymous referee for their useful comments and suggestions that helped in enhancing the contents of this study. AAK is supported by an appointment to the NASA Postdoctoral Program at the Goddard Space Flight Center, administered by the Universities Space Research Association through a contract with NASA. The LAGER survey has been supported in part by the US National Science Foundation through NSF grant AST-1518057. We also thank NASA for financial support via WFIRST Preparatory Science Grant NNX15AJ79G and WFIRST Science Investigation Team contract NNG16PJ33C. LFB was partially supported by CONICYT Project BASAL AFB-17000. CJ acknowledges support from the CAS Key Research Program of Frontier Sciences (No. QYZDB-SSW-SYS033). JW thanks support from NSFC 11421303 \& 11890693. ZYZ is sponsored by Shanghai Pujiang Program, the National Science Foundation of China (11773051), and the CAS Pioneer Hundred Talents Program.

This research has made use of the $z$COSMOS database, operated at CeSAM/LAM, Marseille, France.

This project used data obtained with the Dark Energy Camera (DECam), which was constructed by the Dark Energy Survey (DES) collaboration. Funding for the DES Projects has been provided by the U.S. Department of Energy, the U.S. National Science Foundation, the Ministry of Science and Education of Spain, the Science and Technology Facilities Council of the United Kingdom, the Higher Education Funding Council for England, the National Center for Supercomputing Applications at the University of Illinois at Urbana-Champaign, the Kavli Institute of Cosmological Physics at the University of Chicago, Center for Cosmology and Astro-Particle Physics at the Ohio State University, the Mitchell Institute for Fundamental Physics and Astronomy at Texas A\&M University, Financiadora de Estudos e Projetos, Funda\c{c}\~{a}o Carlos Chagas Filho de Amparo, Financiadora de Estudos e Projetos, Funda\c{c}\~{a}o Carlos Chagas Filho de Amparo \`{a} Pesquisa do Estado do Rio de Janeiro, Conselho Nacional de Desenvolvimento Cient\'{i}fico e Tecnol\'{o}gico and the Minist\'{e}rio da Ci\^{e}ncia, Tecnologia e Inova\c{c}\~{a}o, the Deutsche Forschungsgemeinschaft and the Collaborating Institutions in the Dark Energy Survey. 

The Collaborating Institutions are Argonne National Laboratory, the University of California at Santa Cruz, the University of Cambridge, Centro de Investigaciones En\'{e}rgeticas, Medioambientales y Tecnol\'{o}gicas-Madrid, the University of Chicago, University College London, the DES-Brazil Consortium, the University of Edinburgh, the Eidgen\"{o}ssische Technische Hochschule (ETH) Z\"{u}rich, Fermi National Accelerator Laboratory, the University of Illinois at Urbana-Champaign, the Institut de Ci\`{e}ncies de l'Espai (IEEC/CSIC), the Institut de F\'{i}sica d'Altes Energies, Lawrence Berkeley National Laboratory, the Ludwig-Maximilians Universit\"{a}t M\"{u}nchen and the associated Excellence Cluster Universe, the University of Michigan, the National Optical Astronomy Observatory, the University of Nottingham, the Ohio State University, the OzDES Membership Consortium, the University of Pennsylvania, the University of Portsmouth, SLAC National Accelerator Laboratory, Stanford University, the University of Sussex, and Texas A\&M University. 

Based on observations at Cerro Tololo Inter-American Observatory, National Optical Astronomy Observatory (NOAO 2017A-0366, 2017B-0330, 2018A-0371, 2018B-0327; PI: S.~Malhotra; NOAO 2019B-1008, PI: L.~F.~Barrientos), which is operated by the Association of Universities for Research in Astronomy (AURA) under a cooperative agreement with the National Science Foundation. 

\bibliography{LAGER_LFs}

\begin{thebibliography}{}
\makeatletter
\relax
\def\mn@urlcharsother{\let\do\@makeother \do\$\do\&\do\#\do\^\do\_\do\%\do\~}
\def\mn@doi{\begingroup\mn@urlcharsother \@ifnextchar [ {\mn@doi@}
  {\mn@doi@[]}}
\def\mn@doi@[#1]#2{\def\@tempa{#1}\ifx\@tempa\@empty \href
  {http://dx.doi.org/#2} {doi:#2}\else \href {http://dx.doi.org/#2} {#1}\fi
  \endgroup}
\def\mn@eprint#1#2{\mn@eprint@#1:#2::\@nil}
\def\mn@eprint@arXiv#1{\href {http://arxiv.org/abs/#1} {{\tt arXiv:#1}}}
\def\mn@eprint@dblp#1{\href {http://dblp.uni-trier.de/rec/bibtex/#1.xml}
  {dblp:#1}}
\def\mn@eprint@#1:#2:#3:#4\@nil{\def\@tempa {#1}\def\@tempb {#2}\def\@tempc
  {#3}\ifx \@tempc \@empty \let \@tempc \@tempb \let \@tempb \@tempa \fi \ifx
  \@tempb \@empty \def\@tempb {arXiv}\fi \@ifundefined
  {mn@eprint@\@tempb}{\@tempb:\@tempc}{\expandafter \expandafter \csname
  mn@eprint@\@tempb\endcsname \expandafter{\@tempc}}}

\bibitem[\protect\citeauthoryear{{Alavi} et~al.,}{{Alavi}
  et~al.}{2014}]{Alavi2014}
{Alavi} A.,  et~al., 2014, \mn@doi [\apj] {10.1088/0004-637X/780/2/143}, \href
  {https://ui.adsabs.harvard.edu/abs/2014ApJ...780..143A} {780, 143}

\bibitem[\protect\citeauthoryear{{Alonso-Herrero} et~al.,}{{Alonso-Herrero}
  et~al.}{2006}]{Alonso2006}
{Alonso-Herrero} A.,  et~al., 2006, \mn@doi [\apj] {10.1086/499800}, \href
  {https://ui.adsabs.harvard.edu/abs/2006ApJ...640..167A} {640, 167}

\bibitem[\protect\citeauthoryear{{Azadi} et~al.,}{{Azadi}
  et~al.}{2017}]{Azadi2017}
{Azadi} M.,  et~al., 2017, \mn@doi [\apj] {10.3847/1538-4357/835/1/27}, \href
  {https://ui.adsabs.harvard.edu/abs/2017ApJ...835...27A} {835, 27}

\bibitem[\protect\citeauthoryear{{Baldwin}, {Phillips}  \&
  {Terlevich}}{{Baldwin} et~al.}{1981}]{Baldwin1981}
{Baldwin} J.~A.,  {Phillips} M.~M.,   {Terlevich} R.,  1981, \mn@doi [\pasp]
  {10.1086/130766}, \href
  {https://ui.adsabs.harvard.edu/abs/1981PASP...93....5B} {93, 5}

\bibitem[\protect\citeauthoryear{{Balogh} et~al.,}{{Balogh}
  et~al.}{2014}]{Balogh2014}
{Balogh} M.~L.,  et~al., 2014, \mn@doi [\mnras] {10.1093/mnras/stu1332}, \href
  {https://ui.adsabs.harvard.edu/\#abs/2014MNRAS.443.2679B} {443, 2679}

\bibitem[\protect\citeauthoryear{{Bayliss}, {McMahon}, {Venemans}, {Ryan-Weber}
   \& {Lewis}}{{Bayliss} et~al.}{2011}]{Bayliss2011}
{Bayliss} K.~D.,  {McMahon} R.~G.,  {Venemans} B.~P.,  {Ryan-Weber} E.~V.,
  {Lewis} J.~R.,  2011, \mn@doi [\mnras] {10.1111/j.1365-2966.2011.18360.x},
  \href {https://ui.adsabs.harvard.edu/abs/2011MNRAS.413.2883B} {413, 2883}

\bibitem[\protect\citeauthoryear{{Bayliss}, {McMahon}, {Venemans}, {Banerji}
  \& {Lewis}}{{Bayliss} et~al.}{2012}]{Bayliss2012}
{Bayliss} K.~D.,  {McMahon} R.~G.,  {Venemans} B.~P.,  {Banerji} M.,   {Lewis}
  J.~R.,  2012, \mn@doi [\mnras] {10.1111/j.1365-2966.2012.21683.x}, \href
  {https://ui.adsabs.harvard.edu/abs/2012MNRAS.426.2178B} {426, 2178}

\bibitem[\protect\citeauthoryear{{Bertin} \& {Arnouts}}{{Bertin} \&
  {Arnouts}}{1996}]{Bertin1996}
{Bertin} E.,  {Arnouts} S.,  1996, \mn@doi [\aaps] {10.1051/aas:1996164}, \href
  {https://ui.adsabs.harvard.edu/abs/1996A%26AS..117..393B} {117, 393}

\bibitem[\protect\citeauthoryear{{Bongiorno} et~al.,}{{Bongiorno}
  et~al.}{2010}]{Bongiorno2010}
{Bongiorno} A.,  et~al., 2010, \mn@doi [\aap] {10.1051/0004-6361/200913229},
  \href {https://ui.adsabs.harvard.edu/abs/2010A%26A...510A..56B} {510, A56}

\bibitem[\protect\citeauthoryear{{Bouwens} et~al.,}{{Bouwens}
  et~al.}{2015}]{Bouwens2015}
{Bouwens} R.~J.,  et~al., 2015, \mn@doi [\apj] {10.1088/0004-637X/803/1/34},
  \href {https://ui.adsabs.harvard.edu/abs/2015ApJ...803...34B} {803, 34}

\bibitem[\protect\citeauthoryear{{Brammer} et~al.,}{{Brammer}
  et~al.}{2012}]{Brammer2012}
{Brammer} G.~B.,  et~al., 2012, \mn@doi [\apjs] {10.1088/0067-0049/200/2/13},
  \href {http://adsabs.harvard.edu/abs/2012ApJS..200...13B} {200, 13}

\bibitem[\protect\citeauthoryear{{Brusa} et~al.,}{{Brusa}
  et~al.}{2010}]{Brusa2010}
{Brusa} M.,  et~al., 2010, \mn@doi [\apj] {10.1088/0004-637X/716/1/348}, \href
  {https://ui.adsabs.harvard.edu/abs/2010ApJ...716..348B} {716, 348}

\bibitem[\protect\citeauthoryear{{Calzetti}, {Armus}, {Bohlin}, {Kinney},
  {Koornneef}  \& {Storchi-Bergmann}}{{Calzetti} et~al.}{2000}]{Calzetti2000}
{Calzetti} D.,  {Armus} L.,  {Bohlin} R.~C.,  {Kinney} A.~L.,  {Koornneef} J.,
   {Storchi-Bergmann} T.,  2000, \mn@doi [\apj] {10.1086/308692}, \href
  {http://adsabs.harvard.edu/abs/2000ApJ...533..682C} {533, 682}

\bibitem[\protect\citeauthoryear{{Cappelluti} et~al.,}{{Cappelluti}
  et~al.}{2007}]{Cappelluti2007}
{Cappelluti} N.,  et~al., 2007, \mn@doi [\apjs] {10.1086/516586}, \href
  {https://ui.adsabs.harvard.edu/abs/2007ApJS..172..341C} {172, 341}

\bibitem[\protect\citeauthoryear{{Cardelli}, {Clayton}  \& {Mathis}}{{Cardelli}
  et~al.}{1989}]{Cardelli1989}
{Cardelli} J.~A.,  {Clayton} G.~C.,   {Mathis} J.~S.,  1989, \mn@doi [\apj]
  {10.1086/167900}, \href
  {https://ui.adsabs.harvard.edu/abs/1989ApJ...345..245C} {345, 245}

\bibitem[\protect\citeauthoryear{{Casey}, {Narayanan}  \& {Cooray}}{{Casey}
  et~al.}{2014}]{Casey2014}
{Casey} C.~M.,  {Narayanan} D.,   {Cooray} A.,  2014, \mn@doi [\physrep]
  {10.1016/j.physrep.2014.02.009}, \href
  {https://ui.adsabs.harvard.edu/abs/2014PhR...541...45C} {541, 45}

\bibitem[\protect\citeauthoryear{{Ciardullo} et~al.,}{{Ciardullo}
  et~al.}{2013}]{Ciardullo2013}
{Ciardullo} R.,  et~al., 2013, \mn@doi [\apj] {10.1088/0004-637X/769/1/83},
  \href {https://ui.adsabs.harvard.edu/abs/2013ApJ...769...83C} {769, 83}

\bibitem[\protect\citeauthoryear{{Civano} et~al.,}{{Civano}
  et~al.}{2012}]{Civano2012}
{Civano} F.,  et~al., 2012, \mn@doi [\apjs] {10.1088/0067-0049/201/2/30}, \href
  {https://ui.adsabs.harvard.edu/abs/2012ApJS..201...30C} {201, 30}

\bibitem[\protect\citeauthoryear{{Coil} et~al.,}{{Coil}
  et~al.}{2011}]{Coil2011}
{Coil} A.~L.,  et~al., 2011, \mn@doi [\apj] {10.1088/0004-637X/741/1/8}, \href
  {http://adsabs.harvard.edu/abs/2011ApJ...741....8C} {741, 8}

\bibitem[\protect\citeauthoryear{{Colbert} et~al.,}{{Colbert}
  et~al.}{2013}]{Colbert2013}
{Colbert} J.~W.,  et~al., 2013, \mn@doi [\apj] {10.1088/0004-637X/779/1/34},
  \href {https://ui.adsabs.harvard.edu/abs/2013ApJ...779...34C} {779, 34}

\bibitem[\protect\citeauthoryear{{Comparat} et~al.,}{{Comparat}
  et~al.}{2015}]{Comparat2015}
{Comparat} J.,  et~al., 2015, \mn@doi [\aap] {10.1051/0004-6361/201424767},
  \href {http://adsabs.harvard.edu/abs/2015A%26A...575A..40C} {575, A40}

\bibitem[\protect\citeauthoryear{{Cool} et~al.,}{{Cool}
  et~al.}{2013}]{Cool2013}
{Cool} R.~J.,  et~al., 2013, \mn@doi [\apj] {10.1088/0004-637X/767/2/118},
  \href {http://adsabs.harvard.edu/abs/2013ApJ...767..118C} {767, 118}

\bibitem[\protect\citeauthoryear{{Coughlin} et~al.,}{{Coughlin}
  et~al.}{2018}]{Coughlin2018}
{Coughlin} A.,  et~al., 2018, \mn@doi [\apj] {10.3847/1538-4357/aab620}, \href
  {https://ui.adsabs.harvard.edu/abs/2018ApJ...858...96C} {858, 96}

\bibitem[\protect\citeauthoryear{{Cucciati} et~al.,}{{Cucciati}
  et~al.}{2012}]{Cucciati2012}
{Cucciati} O.,  et~al., 2012, \mn@doi [\aap] {10.1051/0004-6361/201118010},
  \href {https://ui.adsabs.harvard.edu/abs/2012A&A...539A..31C} {539, A31}

\bibitem[\protect\citeauthoryear{{Darvish}, {Mobasher}, {Martin}, {Sobral},
  {Scoville}, {Stroe}, {Hemmati}  \& {Kartaltepe}}{{Darvish}
  et~al.}{2017}]{Darvish2017}
{Darvish} B.,  {Mobasher} B.,  {Martin} D.~C.,  {Sobral} D.,  {Scoville} N.,
  {Stroe} A.,  {Hemmati} S.,   {Kartaltepe} J.,  2017, \mn@doi [\apj]
  {10.3847/1538-4357/837/1/16}, \href
  {https://ui.adsabs.harvard.edu/abs/2017ApJ...837...16D} {837, 16}

\bibitem[\protect\citeauthoryear{{Donley}, {Rieke}, {P{\'e}rez-Gonz{\'a}lez},
  {Rigby}  \& {Alonso-Herrero}}{{Donley} et~al.}{2007}]{Donley2007}
{Donley} J.~L.,  {Rieke} G.~H.,  {P{\'e}rez-Gonz{\'a}lez} P.~G.,  {Rigby}
  J.~R.,   {Alonso-Herrero} A.,  2007, \mn@doi [\apj] {10.1086/512798}, \href
  {https://ui.adsabs.harvard.edu/abs/2007ApJ...660..167D} {660, 167}

\bibitem[\protect\citeauthoryear{{Donley} et~al.,}{{Donley}
  et~al.}{2012}]{Donley2012}
{Donley} J.~L.,  et~al., 2012, \mn@doi [\apj] {10.1088/0004-637X/748/2/142},
  \href {https://ui.adsabs.harvard.edu/abs/2012ApJ...748..142D} {748, 142}

\bibitem[\protect\citeauthoryear{{Elvis} et~al.,}{{Elvis}
  et~al.}{2009}]{Elvis2009}
{Elvis} M.,  et~al., 2009, \mn@doi [\apjs] {10.1088/0067-0049/184/1/158}, \href
  {https://ui.adsabs.harvard.edu/abs/2009ApJS..184..158E} {184, 158}

\bibitem[\protect\citeauthoryear{{Faisst}, {Masters}, {Wang}, {Merson},
  {Capak}, {Malhotra}  \& {Rhoads}}{{Faisst} et~al.}{2018}]{Faisst2018}
{Faisst} A.~L.,  {Masters} D.,  {Wang} Y.,  {Merson} A.,  {Capak} P.,
  {Malhotra} S.,   {Rhoads} J.~E.,  2018, \mn@doi [\apj]
  {10.3847/1538-4357/aab1fc}, \href
  {https://ui.adsabs.harvard.edu/abs/2018ApJ...855..132F} {855, 132}

\bibitem[\protect\citeauthoryear{{Fujita} et~al.,}{{Fujita}
  et~al.}{2003}]{Fujita2003}
{Fujita} S.~S.,  et~al., 2003, \mn@doi [\apj] {10.1086/374859}, \href
  {https://ui.adsabs.harvard.edu/abs/2003ApJ...586L.115F} {586, L115}

\bibitem[\protect\citeauthoryear{{Gallego}, {Zamorano}, {Aragon-Salamanca}  \&
  {Rego}}{{Gallego} et~al.}{1995}]{Gallego1995}
{Gallego} J.,  {Zamorano} J.,  {Aragon-Salamanca} A.,   {Rego} M.,  1995,
  \mn@doi [\apj] {10.1086/309804}, \href
  {https://ui.adsabs.harvard.edu/abs/1995ApJ...455L...1G} {455, L1}

\bibitem[\protect\citeauthoryear{{Gallego}, {Zamorano}, {Rego}  \&
  {Vitores}}{{Gallego} et~al.}{1997}]{Gallego1997}
{Gallego} J.,  {Zamorano} J.,  {Rego} M.,   {Vitores} A.~G.,  1997, \mn@doi
  [\apj] {10.1086/303551}, \href
  {https://ui.adsabs.harvard.edu/abs/1997ApJ...475..502G} {475, 502}

\bibitem[\protect\citeauthoryear{{Garn} \& {Best}}{{Garn} \&
  {Best}}{2010}]{Garn2010}
{Garn} T.,  {Best} P.~N.,  2010, \mn@doi [\mnras]
  {10.1111/j.1365-2966.2010.17321.x}, \href
  {https://ui.adsabs.harvard.edu/abs/2010MNRAS.409..421G} {409, 421}

\bibitem[\protect\citeauthoryear{{Grogin} et~al.,}{{Grogin}
  et~al.}{2011}]{Grogin2011}
{Grogin} N.~A.,  et~al., 2011, \mn@doi [\apjs] {10.1088/0067-0049/197/2/35},
  \href {https://ui.adsabs.harvard.edu/abs/2011ApJS..197...35G} {197, 35}

\bibitem[\protect\citeauthoryear{{Groves}, {Brinchmann}  \& {Walcher}}{{Groves}
  et~al.}{2012}]{Groves2012}
{Groves} B.,  {Brinchmann} J.,   {Walcher} C.~J.,  2012, \mn@doi [\mnras]
  {10.1111/j.1365-2966.2011.19796.x}, \href
  {https://ui.adsabs.harvard.edu/abs/2012MNRAS.419.1402G} {419, 1402}

\bibitem[\protect\citeauthoryear{{Gruppioni} et~al.,}{{Gruppioni}
  et~al.}{2013}]{Gruppioni2013}
{Gruppioni} C.,  et~al., 2013, \mn@doi [\mnras] {10.1093/mnras/stt308}, \href
  {https://ui.adsabs.harvard.edu/abs/2013MNRAS.432...23G} {432, 23}

\bibitem[\protect\citeauthoryear{{Hasinger} et~al.,}{{Hasinger}
  et~al.}{2007}]{Hasinger2007}
{Hasinger} G.,  et~al., 2007, \mn@doi [\apjs] {10.1086/516576}, \href
  {https://ui.adsabs.harvard.edu/abs/2007ApJS..172...29H} {172, 29}

\bibitem[\protect\citeauthoryear{{Hasinger} et~al.,}{{Hasinger}
  et~al.}{2018}]{Hasinger2018}
{Hasinger} G.,  et~al., 2018, \mn@doi [\apj] {10.3847/1538-4357/aabacf}, \href
  {http://adsabs.harvard.edu/abs/2018ApJ...858...77H} {858, 77}

\bibitem[\protect\citeauthoryear{{Hayashi}, {Sobral}, {Best}, {Smail}  \&
  {Kodama}}{{Hayashi} et~al.}{2013}]{Hayashi2013}
{Hayashi} M.,  {Sobral} D.,  {Best} P.~N.,  {Smail} I.,   {Kodama} T.,  2013,
  \mn@doi [\mnras] {10.1093/mnras/sts676}, \href
  {https://ui.adsabs.harvard.edu/abs/2013MNRAS.430.1042H} {430, 1042}

\bibitem[\protect\citeauthoryear{{Hayashi} et~al.,}{{Hayashi}
  et~al.}{2015}]{Hayashi2015}
{Hayashi} M.,  et~al., 2015, \mn@doi [\pasj] {10.1093/pasj/psv041}, \href
  {https://ui.adsabs.harvard.edu/abs/2015PASJ...67...80H} {67, 80}

\bibitem[\protect\citeauthoryear{{Hayashi} et~al.,}{{Hayashi}
  et~al.}{2018}]{Hayashi2018}
{Hayashi} M.,  et~al., 2018, \mn@doi [\pasj] {10.1093/pasj/psx088}, \href
  {https://ui.adsabs.harvard.edu/abs/2018PASJ...70S..17H} {70, S17}

\bibitem[\protect\citeauthoryear{{Heckman}, {Ptak}, {Hornschemeier}  \&
  {Kauffmann}}{{Heckman} et~al.}{2005}]{Heckman2005}
{Heckman} T.~M.,  {Ptak} A.,  {Hornschemeier} A.,   {Kauffmann} G.,  2005,
  \mn@doi [\apj] {10.1086/491665}, \href
  {https://ui.adsabs.harvard.edu/abs/2005ApJ...634..161H} {634, 161}

\bibitem[\protect\citeauthoryear{{Hippelein} et~al.,}{{Hippelein}
  et~al.}{2003}]{Hippelein2003}
{Hippelein} H.,  et~al., 2003, \mn@doi [\aap] {10.1051/0004-6361:20021898},
  \href {https://ui.adsabs.harvard.edu/abs/2003A&A...402...65H} {402, 65}

\bibitem[\protect\citeauthoryear{{Hopkins}}{{Hopkins}}{2004}]{Hopkins2004}
{Hopkins} A.~M.,  2004, \mn@doi [\apj] {10.1086/424032}, \href
  {https://ui.adsabs.harvard.edu/abs/2004ApJ...615..209H} {615, 209}

\bibitem[\protect\citeauthoryear{{Hopkins} \& {Beacom}}{{Hopkins} \&
  {Beacom}}{2006}]{Hopkins2006}
{Hopkins} A.~M.,  {Beacom} J.~F.,  2006, \mn@doi [\apj] {10.1086/506610}, \href
  {https://ui.adsabs.harvard.edu/abs/2006ApJ...651..142H} {651, 142}

\bibitem[\protect\citeauthoryear{{Hopkins}, {Connolly}, {Haarsma}  \&
  {Cram}}{{Hopkins} et~al.}{2001}]{Hopkins2001}
{Hopkins} A.~M.,  {Connolly} A.~J.,  {Haarsma} D.~B.,   {Cram} L.~E.,  2001,
  \mn@doi [\aj] {10.1086/321113}, \href
  {https://ui.adsabs.harvard.edu/abs/2001AJ....122..288H} {122, 288}

\bibitem[\protect\citeauthoryear{{Hu} et~al.,}{{Hu} et~al.}{2017}]{Hu2017}
{Hu} W.,  et~al., 2017, \mn@doi [\apjl] {10.3847/2041-8213/aa8401}, \href
  {https://ui.adsabs.harvard.edu/abs/2017ApJ...845L..16H} {845, L16}

\bibitem[\protect\citeauthoryear{{Hu} et~al.,}{{Hu} et~al.}{2019}]{Hu2019}
{Hu} W.,  et~al., 2019, arXiv e-prints, \href
  {https://ui.adsabs.harvard.edu/abs/2019arXiv190309046H} {p. arXiv:1903.09046}

\bibitem[\protect\citeauthoryear{{Ibar} et~al.,}{{Ibar}
  et~al.}{2013}]{Ibar2013}
{Ibar} E.,  et~al., 2013, \mn@doi [\mnras] {10.1093/mnras/stt1258}, \href
  {https://ui.adsabs.harvard.edu/abs/2013MNRAS.434.3218I} {434, 3218}

\bibitem[\protect\citeauthoryear{{Juneau}, {Dickinson}, {Alexander}  \&
  {Salim}}{{Juneau} et~al.}{2011}]{Juneau2011}
{Juneau} S.,  {Dickinson} M.,  {Alexander} D.~M.,   {Salim} S.,  2011, \mn@doi
  [\apj] {10.1088/0004-637X/736/2/104}, \href
  {https://ui.adsabs.harvard.edu/abs/2011ApJ...736..104J} {736, 104}

\bibitem[\protect\citeauthoryear{{Kaasinen}, {Bian}, {Groves}, {Kewley}  \&
  {Gupta}}{{Kaasinen} et~al.}{2017}]{Kaasinen2017}
{Kaasinen} M.,  {Bian} F.,  {Groves} B.,  {Kewley} L.~J.,   {Gupta} A.,  2017,
  \mn@doi [\mnras] {10.1093/mnras/stw2827}, \href
  {http://adsabs.harvard.edu/abs/2017MNRAS.465.3220K} {465, 3220}

\bibitem[\protect\citeauthoryear{{Karim} et~al.,}{{Karim}
  et~al.}{2011}]{Karim2011}
{Karim} A.,  et~al., 2011, \mn@doi [\apj] {10.1088/0004-637X/730/2/61}, \href
  {https://ui.adsabs.harvard.edu/abs/2011ApJ...730...61K} {730, 61}

\bibitem[\protect\citeauthoryear{{Kauffmann} \& {Heckman}}{{Kauffmann} \&
  {Heckman}}{2009}]{Kauffmann2009}
{Kauffmann} G.,  {Heckman} T.~M.,  2009, \mn@doi [\mnras]
  {10.1111/j.1365-2966.2009.14960.x}, \href
  {https://ui.adsabs.harvard.edu/abs/2009MNRAS.397..135K} {397, 135}

\bibitem[\protect\citeauthoryear{{Kauffmann} et~al.,}{{Kauffmann}
  et~al.}{2003}]{Kauffmann2003}
{Kauffmann} G.,  et~al., 2003, \mn@doi [\mnras]
  {10.1111/j.1365-2966.2003.07154.x}, \href
  {https://ui.adsabs.harvard.edu/abs/2003MNRAS.346.1055K} {346, 1055}

\bibitem[\protect\citeauthoryear{{Kennicutt}}{{Kennicutt}}{1992}]{Kennicutt1992}
{Kennicutt} Robert~C. J.,  1992, \mn@doi [\apj] {10.1086/171154}, \href
  {https://ui.adsabs.harvard.edu/abs/1992ApJ...388..310K} {388, 310}

\bibitem[\protect\citeauthoryear{{Kennicutt}}{{Kennicutt}}{1998}]{Kennicutt1998}
{Kennicutt} Jr. R.~C.,  1998, \mn@doi [\araa] {10.1146/annurev.astro.36.1.189},
  \href {https://ui.adsabs.harvard.edu/abs/1998ARA%26A..36..189K} {36, 189}

\bibitem[\protect\citeauthoryear{{Khostovan}, {Sobral}, {Mobasher}, {Best},
  {Smail}, {Stott}, {Hemmati}  \& {Nayyeri}}{{Khostovan}
  et~al.}{2015}]{Khostovan2015}
{Khostovan} A.~A.,  {Sobral} D.,  {Mobasher} B.,  {Best} P.~N.,  {Smail} I.,
  {Stott} J.~P.,  {Hemmati} S.,   {Nayyeri} H.,  2015, \mn@doi [\mnras]
  {10.1093/mnras/stv1474}, \href
  {http://adsabs.harvard.edu/abs/2015MNRAS.452.3948K} {452, 3948}

\bibitem[\protect\citeauthoryear{{Khostovan}, {Sobral}, {Mobasher}, {Smail},
  {Darvish}, {Nayyeri}, {Hemmati}  \& {Stott}}{{Khostovan}
  et~al.}{2016}]{Khostovan2016}
{Khostovan} A.~A.,  {Sobral} D.,  {Mobasher} B.,  {Smail} I.,  {Darvish} B.,
  {Nayyeri} H.,  {Hemmati} S.,   {Stott} J.~P.,  2016, \mn@doi [\mnras]
  {10.1093/mnras/stw2174}, \href
  {http://adsabs.harvard.edu/abs/2016MNRAS.463.2363K} {463, 2363}

\bibitem[\protect\citeauthoryear{{Koekemoer} et~al.,}{{Koekemoer}
  et~al.}{2007}]{Koekemoer2007}
{Koekemoer} A.~M.,  et~al., 2007, \mn@doi [\apjs] {10.1086/520086}, \href
  {https://ui.adsabs.harvard.edu/abs/2007ApJS..172..196K} {172, 196}

\bibitem[\protect\citeauthoryear{{Kriek} et~al.,}{{Kriek}
  et~al.}{2015}]{Kriek2015}
{Kriek} M.,  et~al., 2015, \mn@doi [\apjs] {10.1088/0067-0049/218/2/15}, \href
  {http://adsabs.harvard.edu/abs/2015ApJS..218...15K} {218, 15}

\bibitem[\protect\citeauthoryear{{Laigle} et~al.,}{{Laigle}
  et~al.}{2016}]{Laigle2016}
{Laigle} C.,  et~al., 2016, \mn@doi [\apjs] {10.3847/0067-0049/224/2/24}, \href
  {http://adsabs.harvard.edu/abs/2016ApJS..224...24L} {224, 24}

\bibitem[\protect\citeauthoryear{{Lilly} et~al.,}{{Lilly}
  et~al.}{2009}]{Lilly2009}
{Lilly} S.~J.,  et~al., 2009, \mn@doi [\apjs] {10.1088/0067-0049/184/2/218},
  \href {http://adsabs.harvard.edu/abs/2009ApJS..184..218L} {184, 218}

\bibitem[\protect\citeauthoryear{{Ly} et~al.,}{{Ly} et~al.}{2007}]{Ly2007}
{Ly} C.,  et~al., 2007, \mn@doi [\apj] {10.1086/510828}, \href
  {http://adsabs.harvard.edu/abs/2007ApJ...657..738L} {657, 738}

\bibitem[\protect\citeauthoryear{{Ly}, {Lee}, {Dale}, {Momcheva}, {Salim},
  {Staudaher}, {Moore}  \& {Finn}}{{Ly} et~al.}{2011}]{Ly2011}
{Ly} C.,  {Lee} J.~C.,  {Dale} D.~A.,  {Momcheva} I.,  {Salim} S.,  {Staudaher}
  S.,  {Moore} C.~A.,   {Finn} R.,  2011, \mn@doi [\apj]
  {10.1088/0004-637X/726/2/109}, \href
  {https://ui.adsabs.harvard.edu/abs/2011ApJ...726..109L} {726, 109}

\bibitem[\protect\citeauthoryear{{Madau} \& {Dickinson}}{{Madau} \&
  {Dickinson}}{2014}]{Madau2014}
{Madau} P.,  {Dickinson} M.,  2014, \mn@doi [\araa]
  {10.1146/annurev-astro-081811-125615}, \href
  {https://ui.adsabs.harvard.edu/abs/2014ARA%26A..52..415M} {52, 415}

\bibitem[\protect\citeauthoryear{{Magnelli}, {Elbaz}, {Chary}, {Dickinson}, {Le
  Borgne}, {Frayer}  \& {Willmer}}{{Magnelli} et~al.}{2011}]{Magnelli2011}
{Magnelli} B.,  {Elbaz} D.,  {Chary} R.~R.,  {Dickinson} M.,  {Le Borgne} D.,
  {Frayer} D.~T.,   {Willmer} C.~N.~A.,  2011, \mn@doi [\aap]
  {10.1051/0004-6361/200913941}, \href
  {https://ui.adsabs.harvard.edu/abs/2011A&A...528A..35M} {528, A35}

\bibitem[\protect\citeauthoryear{{Magnelli} et~al.,}{{Magnelli}
  et~al.}{2013}]{Magnelli2013}
{Magnelli} B.,  et~al., 2013, \mn@doi [\aap] {10.1051/0004-6361/201321371},
  \href {https://ui.adsabs.harvard.edu/abs/2013A&A...553A.132M} {553, A132}

\bibitem[\protect\citeauthoryear{{Masters}, {Stern}, {Cohen}, {Capak},
  {Rhodes}, {Castander}  \& {Paltani}}{{Masters} et~al.}{2017}]{Masters2017}
{Masters} D.~C.,  {Stern} D.~K.,  {Cohen} J.~G.,  {Capak} P.~L.,  {Rhodes}
  J.~D.,  {Castander} F.~J.,   {Paltani} S.,  2017, \mn@doi [\apj]
  {10.3847/1538-4357/aa6f08}, \href
  {http://adsabs.harvard.edu/abs/2017ApJ...841..111M} {841, 111}

\bibitem[\protect\citeauthoryear{{Matthee}, {Sobral}, {Best}, {Smail}, {Bian},
  {Darvish}, {R{\"o}ttgering}  \& {Fan}}{{Matthee} et~al.}{2017}]{Matthee2017}
{Matthee} J.,  {Sobral} D.,  {Best} P.,  {Smail} I.,  {Bian} F.,  {Darvish} B.,
   {R{\"o}ttgering} H.,   {Fan} X.,  2017, \mn@doi [\mnras]
  {10.1093/mnras/stx1569}, \href
  {http://adsabs.harvard.edu/abs/2017MNRAS.471..629M} {471, 629}

\bibitem[\protect\citeauthoryear{{McCracken} et~al.,}{{McCracken}
  et~al.}{2010}]{McCracken2010}
{McCracken} H.~J.,  et~al., 2010, \mn@doi [\apj] {10.1088/0004-637X/708/1/202},
  \href {https://ui.adsabs.harvard.edu/abs/2010ApJ...708..202M} {708, 202}

\bibitem[\protect\citeauthoryear{{McCracken} et~al.,}{{McCracken}
  et~al.}{2012}]{McCracken2012}
{McCracken} H.~J.,  et~al., 2012, \mn@doi [\aap] {10.1051/0004-6361/201219507},
  \href {https://ui.adsabs.harvard.edu/abs/2012A&A...544A.156M} {544, A156}

\bibitem[\protect\citeauthoryear{{Merson}, {Wang}, {Benson}, {Faisst},
  {Masters}, {Kiessling}  \& {Rhodes}}{{Merson} et~al.}{2018}]{Merson2018}
{Merson} A.,  {Wang} Y.,  {Benson} A.,  {Faisst} A.,  {Masters} D.,
  {Kiessling} A.,   {Rhodes} J.,  2018, \mn@doi [\mnras]
  {10.1093/mnras/stx2649}, \href
  {https://ui.adsabs.harvard.edu/abs/2018MNRAS.474..177M} {474, 177}

\bibitem[\protect\citeauthoryear{{Momcheva}, {Lee}, {Ly}, {Salim}, {Dale},
  {Ouchi}, {Finn}  \& {Ono}}{{Momcheva} et~al.}{2013}]{Momcheva2013}
{Momcheva} I.~G.,  {Lee} J.~C.,  {Ly} C.,  {Salim} S.,  {Dale} D.~A.,  {Ouchi}
  M.,  {Finn} R.,   {Ono} Y.,  2013, \mn@doi [\aj]
  {10.1088/0004-6256/145/2/47}, \href
  {https://ui.adsabs.harvard.edu/abs/2013AJ....145...47M} {145, 47}

\bibitem[\protect\citeauthoryear{{Momcheva} et~al.,}{{Momcheva}
  et~al.}{2016}]{Momcheva2016}
{Momcheva} I.~G.,  et~al., 2016, \mn@doi [\apjs] {10.3847/0067-0049/225/2/27},
  \href {http://adsabs.harvard.edu/abs/2016ApJS..225...27M} {225, 27}

\bibitem[\protect\citeauthoryear{{Morioka}, {Nakajima}, {Taniguchi}, {Shioya},
  {Murayama}  \& {Sasaki}}{{Morioka} et~al.}{2008}]{Morioka2008}
{Morioka} T.,  {Nakajima} A.,  {Taniguchi} Y.,  {Shioya} Y.,  {Murayama} T.,
  {Sasaki} S.~S.,  2008, \mn@doi [\pasj] {10.1093/pasj/60.6.1219}, \href
  {http://adsabs.harvard.edu/abs/2008PASJ...60.1219M} {60, 1219}

\bibitem[\protect\citeauthoryear{{Moustakas}, {Kennicutt}  \&
  {Tremonti}}{{Moustakas} et~al.}{2006}]{Moustakas2006}
{Moustakas} J.,  {Kennicutt} Robert~C. J.,   {Tremonti} C.~A.,  2006, \mn@doi
  [\apj] {10.1086/500964}, \href
  {https://ui.adsabs.harvard.edu/abs/2006ApJ...642..775M} {642, 775}

\bibitem[\protect\citeauthoryear{{Muzzin} et~al.,}{{Muzzin}
  et~al.}{2013}]{Muzzin2013}
{Muzzin} A.,  et~al., 2013, \mn@doi [\apjs] {10.1088/0067-0049/206/1/8}, \href
  {https://ui.adsabs.harvard.edu/abs/2013ApJS..206....8M} {206, 8}

\bibitem[\protect\citeauthoryear{{Osterbrock}}{{Osterbrock}}{1989}]{Osterbrock1989}
{Osterbrock} D.~E.,  1989, {Astrophysics of gaseous nebulae and active galactic
  nuclei}

\bibitem[\protect\citeauthoryear{{Osterbrock} \& {Ferland}}{{Osterbrock} \&
  {Ferland}}{2006}]{Osterbrock2006}
{Osterbrock} D.~E.,  {Ferland} G.~J.,  2006, {Astrophysics of gaseous nebulae
  and active galactic nuclei}

\bibitem[\protect\citeauthoryear{{Pirzkal} et~al.,}{{Pirzkal}
  et~al.}{2013}]{Pirzkal2013}
{Pirzkal} N.,  et~al., 2013, \mn@doi [\apj] {10.1088/0004-637X/772/1/48}, \href
  {https://ui.adsabs.harvard.edu/abs/2013ApJ...772...48P} {772, 48}

\bibitem[\protect\citeauthoryear{{Pirzkal} et~al.,}{{Pirzkal}
  et~al.}{2018}]{Pirzkal2018}
{Pirzkal} N.,  et~al., 2018, \mn@doi [\apj] {10.3847/1538-4357/aae585}, \href
  {https://ui.adsabs.harvard.edu/abs/2018ApJ...868...61P} {868, 61}

\bibitem[\protect\citeauthoryear{{Pozzetti} et~al.,}{{Pozzetti}
  et~al.}{2016}]{Pozzetti2016}
{Pozzetti} L.,  et~al., 2016, \mn@doi [\aap] {10.1051/0004-6361/201527081},
  \href {https://ui.adsabs.harvard.edu/abs/2016A&A...590A...3P} {590, A3}

\bibitem[\protect\citeauthoryear{{Prescott}, {Impey}, {Cool}  \&
  {Scoville}}{{Prescott} et~al.}{2006}]{Prescott2006}
{Prescott} M.~K.~M.,  {Impey} C.~D.,  {Cool} R.~J.,   {Scoville} N.~Z.,  2006,
  \mn@doi [\apj] {10.1086/503325}, \href
  {http://adsabs.harvard.edu/abs/2006ApJ...644..100P} {644, 100}

\bibitem[\protect\citeauthoryear{{Reddy} \& {Steidel}}{{Reddy} \&
  {Steidel}}{2009}]{Reddy2009}
{Reddy} N.~A.,  {Steidel} C.~C.,  2009, \mn@doi [\apj]
  {10.1088/0004-637X/692/1/778}, \href
  {https://ui.adsabs.harvard.edu/abs/2009ApJ...692..778R} {692, 778}

\bibitem[\protect\citeauthoryear{{Sawicki}}{{Sawicki}}{2002}]{Sawicki2002}
{Sawicki} M.,  2002, \mn@doi [\aj] {10.1086/344682}, \href
  {https://ui.adsabs.harvard.edu/abs/2002AJ....124.3050S} {124, 3050}

\bibitem[\protect\citeauthoryear{{Scoville} et~al.,}{{Scoville}
  et~al.}{2007}]{Scoville2007}
{Scoville} N.,  et~al., 2007, \mn@doi [\apjs] {10.1086/516751}, \href
  {https://ui.adsabs.harvard.edu/abs/2007ApJS..172..150S} {172, 150}

\bibitem[\protect\citeauthoryear{{Shioya} et~al.,}{{Shioya}
  et~al.}{2008}]{Shioya2008}
{Shioya} Y.,  et~al., 2008, \mn@doi [\apjs] {10.1086/523703}, \href
  {https://ui.adsabs.harvard.edu/abs/2008ApJS..175..128S} {175, 128}

\bibitem[\protect\citeauthoryear{{Silverman} et~al.,}{{Silverman}
  et~al.}{2015}]{Silverman2015}
{Silverman} J.~D.,  et~al., 2015, \mn@doi [\apjs] {10.1088/0067-0049/220/1/12},
  \href {http://adsabs.harvard.edu/abs/2015ApJS..220...12S} {220, 12}

\bibitem[\protect\citeauthoryear{{Sobral}, {Best}, {Matsuda}, {Smail}, {Geach}
  \& {Cirasuolo}}{{Sobral} et~al.}{2012}]{Sobral2012}
{Sobral} D.,  {Best} P.~N.,  {Matsuda} Y.,  {Smail} I.,  {Geach} J.~E.,
  {Cirasuolo} M.,  2012, \mn@doi [\mnras] {10.1111/j.1365-2966.2011.19977.x},
  \href {http://adsabs.harvard.edu/abs/2012MNRAS.420.1926S} {420, 1926}

\bibitem[\protect\citeauthoryear{{Sobral}, {Smail}, {Best}, {Geach}, {Matsuda},
  {Stott}, {Cirasuolo}  \& {Kurk}}{{Sobral} et~al.}{2013}]{Sobral2013}
{Sobral} D.,  {Smail} I.,  {Best} P.~N.,  {Geach} J.~E.,  {Matsuda} Y.,
  {Stott} J.~P.,  {Cirasuolo} M.,   {Kurk} J.,  2013, \mn@doi [\mnras]
  {10.1093/mnras/sts096}, \href
  {http://adsabs.harvard.edu/abs/2013MNRAS.428.1128S} {428, 1128}

\bibitem[\protect\citeauthoryear{{Sobral} et~al.,}{{Sobral}
  et~al.}{2015}]{Sobral2015}
{Sobral} D.,  et~al., 2015, \mn@doi [\mnras] {10.1093/mnras/stv1076}, \href
  {http://adsabs.harvard.edu/abs/2015MNRAS.451.2303S} {451, 2303}

\bibitem[\protect\citeauthoryear{{Sobral}, {Kohn}, {Best}, {Smail}, {Harrison},
  {Stott}, {Calhau}  \& {Matthee}}{{Sobral} et~al.}{2016a}]{Sobral2016}
{Sobral} D.,  {Kohn} S.~A.,  {Best} P.~N.,  {Smail} I.,  {Harrison} C.~M.,
  {Stott} J.,  {Calhau} J.,   {Matthee} J.,  2016a, \mn@doi [\mnras]
  {10.1093/mnras/stw022}, \href
  {https://ui.adsabs.harvard.edu/abs/2016MNRAS.457.1739S} {457, 1739}

\bibitem[\protect\citeauthoryear{{Sobral}, {Stroe}, {Koyama}, {Darvish},
  {Calhau}, {Afonso}, {Kodama}  \& {Nakata}}{{Sobral}
  et~al.}{2016b}]{Sobral2016b}
{Sobral} D.,  {Stroe} A.,  {Koyama} Y.,  {Darvish} B.,  {Calhau} J.,  {Afonso}
  A.,  {Kodama} T.,   {Nakata} F.,  2016b, \mn@doi [\mnras]
  {10.1093/mnras/stw534}, \href
  {https://ui.adsabs.harvard.edu/abs/2016MNRAS.458.3443S} {458, 3443}

\bibitem[\protect\citeauthoryear{{Sobral}, {Santos}, {Matthee},
  {Paulino-Afonso}, {Ribeiro}, {Calhau}  \& {Khostovan}}{{Sobral}
  et~al.}{2018a}]{Sobral2018b}
{Sobral} D.,  {Santos} S.,  {Matthee} J.,  {Paulino-Afonso} A.,  {Ribeiro} B.,
  {Calhau} J.,   {Khostovan} A.~A.,  2018a, \mn@doi [\mnras]
  {10.1093/mnras/sty378}, \href
  {https://ui.adsabs.harvard.edu/abs/2018MNRAS.476.4725S} {476, 4725}

\bibitem[\protect\citeauthoryear{{Sobral} et~al.,}{{Sobral}
  et~al.}{2018b}]{Sobral2018}
{Sobral} D.,  et~al., 2018b, \mn@doi [\mnras] {10.1093/mnras/sty782}, \href
  {https://ui.adsabs.harvard.edu/abs/2018MNRAS.477.2817S} {477, 2817}

\bibitem[\protect\citeauthoryear{{Steidel} et~al.,}{{Steidel}
  et~al.}{2014}]{Steidel2014}
{Steidel} C.~C.,  et~al., 2014, \mn@doi [\apj] {10.1088/0004-637X/795/2/165},
  \href {https://ui.adsabs.harvard.edu/abs/2014ApJ...795..165S} {795, 165}

\bibitem[\protect\citeauthoryear{{Stern} et~al.,}{{Stern}
  et~al.}{2005}]{Stern2005}
{Stern} D.,  et~al., 2005, \mn@doi [\apj] {10.1086/432523}, \href
  {https://ui.adsabs.harvard.edu/abs/2005ApJ...631..163S} {631, 163}

\bibitem[\protect\citeauthoryear{{Straatman} et~al.,}{{Straatman}
  et~al.}{2018}]{Straatman2018}
{Straatman} C.~M.~S.,  et~al., 2018, \mn@doi [\apjs]
  {10.3847/1538-4365/aae37a}, \href
  {http://adsabs.harvard.edu/abs/2018ApJS..239...27S} {239, 27}

\bibitem[\protect\citeauthoryear{{Stroe} \& {Sobral}}{{Stroe} \&
  {Sobral}}{2015}]{Stroe2015}
{Stroe} A.,  {Sobral} D.,  2015, \mn@doi [\mnras] {10.1093/mnras/stv1555},
  \href {https://ui.adsabs.harvard.edu/abs/2015MNRAS.453..242S} {453, 242}

\bibitem[\protect\citeauthoryear{{Tadaki}, {Kodama}, {Koyama}, {Hayashi},
  {Tanaka}  \& {Tokoku}}{{Tadaki} et~al.}{2011}]{Tadaki2011}
{Tadaki} K.-I.,  {Kodama} T.,  {Koyama} Y.,  {Hayashi} M.,  {Tanaka} I.,
  {Tokoku} C.,  2011, \mn@doi [\pasj] {10.1093/pasj/63.sp2.S437}, \href
  {https://ui.adsabs.harvard.edu/abs/2011PASJ...63S.437T} {63, 437}

\bibitem[\protect\citeauthoryear{{Takahashi} et~al.,}{{Takahashi}
  et~al.}{2007}]{Takahashi2007}
{Takahashi} M.~I.,  et~al., 2007, \mn@doi [\apjs] {10.1086/518037}, \href
  {https://ui.adsabs.harvard.edu/abs/2007ApJS..172..456T} {172, 456}

\bibitem[\protect\citeauthoryear{{Taniguchi} et~al.,}{{Taniguchi}
  et~al.}{2007}]{Taniguchi2007}
{Taniguchi} Y.,  et~al., 2007, \mn@doi [\apjs] {10.1086/516596}, \href
  {https://ui.adsabs.harvard.edu/abs/2007ApJS..172....9T} {172, 9}

\bibitem[\protect\citeauthoryear{{Taniguchi} et~al.,}{{Taniguchi}
  et~al.}{2015}]{Taniguchi2015}
{Taniguchi} Y.,  et~al., 2015, \mn@doi [\pasj] {10.1093/pasj/psv106}, \href
  {https://ui.adsabs.harvard.edu/abs/2015PASJ...67..104T} {67, 104}

\bibitem[\protect\citeauthoryear{{Thomas} et~al.,}{{Thomas}
  et~al.}{2013}]{Thomas2013}
{Thomas} D.,  et~al., 2013, \mn@doi [\mnras] {10.1093/mnras/stt261}, \href
  {https://ui.adsabs.harvard.edu/abs/2013MNRAS.431.1383T} {431, 1383}

\bibitem[\protect\citeauthoryear{{Tresse} \& {Maddox}}{{Tresse} \&
  {Maddox}}{1998}]{Tresse1998}
{Tresse} L.,  {Maddox} S.~J.,  1998, \mn@doi [\apj] {10.1086/305331}, \href
  {https://ui.adsabs.harvard.edu/abs/1998ApJ...495..691T} {495, 691}

\bibitem[\protect\citeauthoryear{{Trump} et~al.,}{{Trump}
  et~al.}{2009}]{Trump2009}
{Trump} J.~R.,  et~al., 2009, \mn@doi [\apj] {10.1088/0004-637X/696/2/1195},
  \href {http://adsabs.harvard.edu/abs/2009ApJ...696.1195T} {696, 1195}

\bibitem[\protect\citeauthoryear{{Valdes}, {Gruendl}  \& {DES
  Project}}{{Valdes} et~al.}{2014}]{Valdes2014}
{Valdes} F.,  {Gruendl} R.,   {DES Project} 2014, in {Manset} N.,  {Forshay}
  P.,  eds,  Astronomical Society of the Pacific Conference Series Vol. 485,
  Astronomical Data Analysis Software and Systems XXIII. p.~379

\bibitem[\protect\citeauthoryear{{Villar}, {Gallego}, {P{\'e}rez-Gonz{\'a}lez},
  {Pascual}, {Noeske}, {Koo}, {Barro}  \& {Zamorano}}{{Villar}
  et~al.}{2008}]{Villar2008}
{Villar} V.,  {Gallego} J.,  {P{\'e}rez-Gonz{\'a}lez} P.~G.,  {Pascual} S.,
  {Noeske} K.,  {Koo} D.~C.,  {Barro} G.,   {Zamorano} J.,  2008, \mn@doi
  [\apj] {10.1086/528942}, \href
  {http://adsabs.harvard.edu/abs/2008ApJ...677..169V} {677, 169}

\bibitem[\protect\citeauthoryear{{Whitaker} et~al.,}{{Whitaker}
  et~al.}{2011}]{Whitaker2011}
{Whitaker} K.~E.,  et~al., 2011, \mn@doi [\apj] {10.1088/0004-637X/735/2/86},
  \href {https://ui.adsabs.harvard.edu/abs/2011ApJ...735...86W} {735, 86}

\bibitem[\protect\citeauthoryear{{Wild}, {Heckman}  \& {Charlot}}{{Wild}
  et~al.}{2010}]{Wild2010}
{Wild} V.,  {Heckman} T.,   {Charlot} S.,  2010, \mn@doi [\mnras]
  {10.1111/j.1365-2966.2010.16536.x}, \href
  {https://ui.adsabs.harvard.edu/abs/2010MNRAS.405..933W} {405, 933}

\bibitem[\protect\citeauthoryear{{Wold}, {Finkelstein}, {Barger}, {Cowie}  \&
  {Rosenwasser}}{{Wold} et~al.}{2017}]{Wold2017}
{Wold} I. G.~B.,  {Finkelstein} S.~L.,  {Barger} A.~J.,  {Cowie} L.~L.,
  {Rosenwasser} B.,  2017, \mn@doi [\apj] {10.3847/1538-4357/aa8d6b}, \href
  {https://ui.adsabs.harvard.edu/abs/2017ApJ...848..108W} {848, 108}

\bibitem[\protect\citeauthoryear{{Yang} et~al.,}{{Yang}
  et~al.}{2019}]{Yang2019}
{Yang} H.,  et~al., 2019, \mn@doi [\apj] {10.3847/1538-4357/ab16ce}, \href
  {https://ui.adsabs.harvard.edu/abs/2019ApJ...876..123Y} {876, 123}

\bibitem[\protect\citeauthoryear{{Zheng} et~al.,}{{Zheng}
  et~al.}{2017}]{Zheng2017}
{Zheng} Z.-Y.,  et~al., 2017, \mn@doi [\apjl] {10.3847/2041-8213/aa794f}, \href
  {https://ui.adsabs.harvard.edu/abs/2017ApJ...842L..22Z} {842, L22}

\bibitem[\protect\citeauthoryear{{Zheng} et~al.,}{{Zheng}
  et~al.}{2019}]{Zheng2019}
{Zheng} Z.-Y.,  et~al., 2019, \mn@doi [\pasp] {10.1088/1538-3873/ab1c32}, \href
  {https://ui.adsabs.harvard.edu/abs/2019PASP..131g4502Z} {131, 074502}

\makeatother
\end{thebibliography}

\bsp	
\label{lastpage}
\end{document}